\documentclass[journal]{vgtc}                
\ifpdf
  \pdfoutput=1\relax                   
  \usepackage{graphicx}                
  \DeclareGraphicsExtensions{.pdf,.png,.jpg,.jpeg} 
\else
  \usepackage{graphicx}                
  \DeclareGraphicsExtensions{.eps}     
\fi%

\graphicspath{{figures/}{pictures/}{images/}{./}} 

\usepackage{microtype}                 
\PassOptionsToPackage{warn}{textcomp}  
\usepackage{textcomp}                  
\usepackage{mathptmx}                  
\usepackage{times}                     
\usepackage{cite}                      
\usepackage{tabu}                      
\usepackage{booktabs}                  

\usepackage[dvipsnames,table]{xcolor}

\onlineid{0}

\vgtccategory{Research}
\vgtcpapertype{please specify}

\title{Visualization for Epidemiological Modelling: \\
\vspace{2px}{Challenges, Solutions, Reflections \& Recommendations}
}

\author{
Jason Dykes$^{1}$,          
Alfie Abdul-Rahman$^{2}$,   
Daniel Archambault$^{3}$,   
Benjamin Bach$^{4}$,        
Rita Borgo$^{2}$,           
\\Min Chen$^{5}$,           
Jessica Enright$^{8}$,      
Hui Fang$^{6}$,             
Elif E. Firat$^{7}$,        
Euan Freeman$^{8}$,         
Tuna Gönen$^{5}$,           
Claire Harris$^{9}$,                    
\\Radu Jianu$^{1}$,         
Nigel W. John$^{10}$,       
Saiful Khan$^{5}$,          
Andrew Lahiff$^{11}$,       
Robert S. Laramee$^{7}$,    
Louise Matthews$^{8}$,      
\\Sibylle Mohr$^{8}$,       
Phong H. Nguyen$^{5}$,      
Alma A. M. Rahat$^{3}$,     
Richard Reeve$^{8}$,        
Panagiotis D. Ritsos$^{12}$,
\\Jonathan C. Roberts$^{12}$, 
Aidan Slingsby$^{1}$,       
Ben Swallow$^{8}$,          
Thomas Torsney-Weir$^{3}$,  
\\Cagatay Turkay$^{13}$,    
Robert Turner$^{14}$,       
Franck P. Vidal$^{12}$,     
Qiru Wang$^{7}$,            
Jo Wood$^{1}$,              
Kai Xu$^{15}$               
}

\authorfooter{
\item
 All authors contributed to the Royal Society's \textit{Rapid Assistance in Modelling the Pandemic: RAMP} initiative.
\item
 \textbf{Jason Dykes} is the lead and corresponding author, and with City, University of London, UK. E-mail: j.dykes@city.ac.uk.
\item
 \textbf{Min Chen} is the submitting author, RAMPVIS project coordinator, and with the University of Oxford, UK. E-mail: min.chen@oerc.ox.ac.uk.
}

\shortauthortitle{Chen \MakeLowercase{\textit{et al.}}: Visualization for Epidemiological Modelling}

\abstract{

We report on an ongoing collaboration between epidemiological modellers and visualization researchers by documenting and reflecting upon knowledge constructs -- a series of ideas, approaches and methods taken from existing visualization research and practice -- deployed and developed to support modelling of the COVID-19 pandemic.
Structured independent commentary on these efforts is synthesized through iterative reflection to develop:
evidence of the effectiveness and value of visualization in this context;
open problems upon which the research communities may focus;
guidance for future activity of this type; and
recommendations
to safeguard the achievements and promote, advance, secure and prepare for future collaborations
of this kind.
In describing and comparing a series of related projects that were undertaken in unprecedented conditions, our hope is that this unique report, and its rich interactive supplementary materials, will guide the scientific community in embracing visualization in its observation, analysis and modelling of data as well as in disseminating findings.
Equally we hope to encourage the visualization community to engage with impactful science in addressing its emerging data challenges.
If we are successful, this showcase of activity may stimulate mutually beneficial engagement between communities with complementary expertise to address problems of significance in epidemiology and beyond.\\
\url{https://ramp-vis.github.io/RAMPVIS-PhilTransA-Supplement/}

} 


\teaser{
  \centering
  \vspace{10px}
  \includegraphics[width=\linewidth]{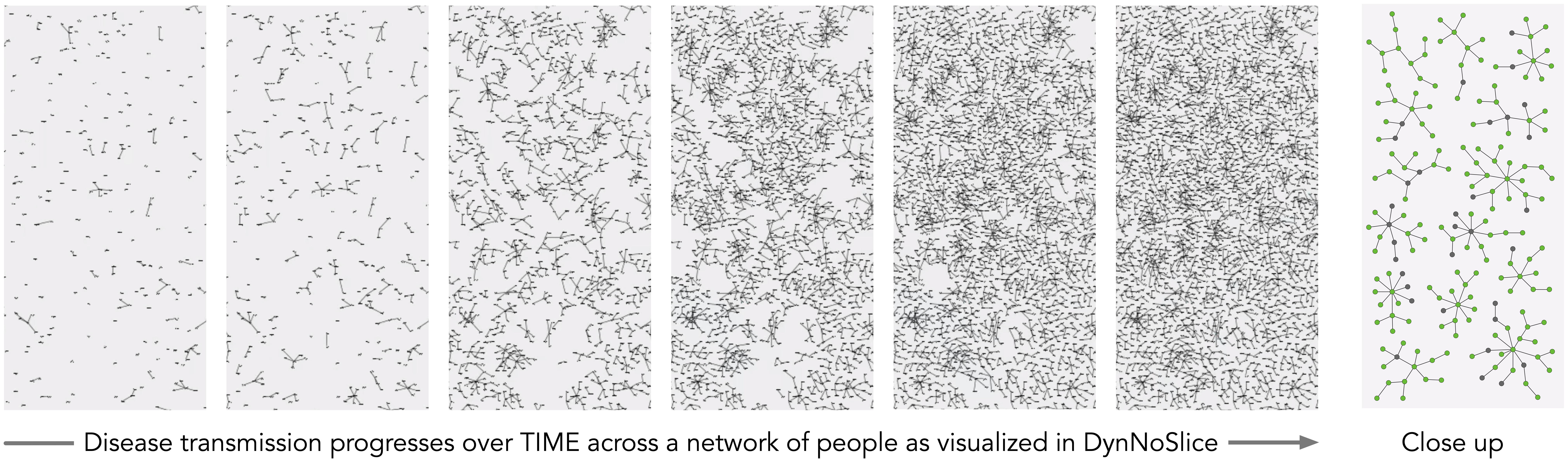}
  \caption{
  Six excerpts of an animation of a modelled disease transmission network \cite{ContactTracing:2020:git} are visualized here using \textit{DynNoSlice} \cite{simonetto2018event, arleo2021multilevel}.
  Points are people, with connections showing infection pathways between them and revealing the fragmented nature of a modelled disease outbreak.
  The final image (right) shows detail, with a close up of the largest components in an infection network, with nodes colour-coded by infection state.
  Modellers responded to this visualization by lowering the random infection rates used in the modelling.
  }
  \label{fig:teaser}
}

\vgtcinsertpkg

\usepackage{dirtytalk}
\usepackage{listofitems}
\usepackage{array}
\usepackage{multibib}
\newcites{obs}{Supplementary Materials:\\~~~~VIS Knowledge Constructs -  Observable Notebooks:}

\newcommand{\summary}[1]{}
\newcommand\rec[1]{
  \begin{itemize}
    \item[$\cdots$] \textbf{#1}
    $\cdots$
    \vspace{4px}
  \end{itemize}
}
\newcommand\effective[1]{
  \begin{itemize}
    \item[$\cdots$] \textit{#1}
    $\cdots$
    \vspace{2px}
  \end{itemize}
}

\newcommand{\secref}[1]{sec.~\ref{#1}}
\definecolor{grey25}{rgb}{0.25, 0.25, 0.25}
\newcommand{\sayIt}[1]{\textcolor{grey25}{\textit{\say{#1}}}}

\newcommand{\etal}{~\textit{et~al.}~}
\newcommand{\etals}{~\textit{et~al's.}~}

\newcommand{\activity}[3]{{\scriptsize\textbf{{#1}-{#2}}} `\textit{#3}'}

\newcommand\numNotebooks{29~}

\newcommand\tinySpace{\hspace{1px}}
\newcommand{\subsubsectionALT}[3]{\subsubsection{#3}}

\newcommand{\checkcell}[4]{
  \ifnum\pdfmatch{#1}{#2}=1
    cell
  \else
    \cite{#2}
  \fi
}

\newcommand\obs[1]{%
\readlist*\token{#1}%
\ifnum\tokenlen=1\relax%
\citeobs{\token[1]}%
\else%
\cell{\token[1]}{\token[2]}%
\fi%
}
\newcommand{\cell}[2]{%
\citeobs{#1}{{\scriptsize{:\textcolor{RoyalBlue}{\underline{\textbf{\href{#1\#cell-#2}{#2}}}}}}}}

\begin{document}






\firstsection{Introduction, Context and Intent}

\maketitle




\summary{
We introduce the discipline(s), the project (RAMP, SCRC, VIS), the nature of the engagement and our organization. We describe the high level structure of what we did - timings, collaborators, etc. We say something about our intention and the body of work that this submission includes. 
VIS is broad, complex and interesting with a deep academic heritage and a vibrant applied research community that allows us to address problems in many areas.
}

Collaboration between epidemiological modellers and visualization researchers 
working in the UK
in response to the COVID-19 pandemic
has been broad, intense and productive.
Research and practice in \textit{visualization and visual analytics} (VIS) have enriched and advanced the models and the modelling process as a result.
The efforts and experiences have in turn contributed valuable knowledge and developments to VIS.
We report and reflect upon this ongoing interdisciplinary activity and the knowledge generated through our collective experience. 





\subsection{Epidemiological Modelling for the COVID-19 Pandemic}
\label{sec:EM}
\label{sub:visInEpMod}
\summary{Epidemiological modelling is difficult and data rich, and contributing to the pandemic.}

Epidemiological modelling has played a significant role in informing policy for the COVID-19 pandemic in the UK \cite{brooks2021modelling, Davies2020, Kucharski2020}
and as evidenced throughout this special issue.
It involves the development and use of mathematical and computational techniques to describe the spread, evolution, and control of epidemic disease \cite{keeling2011modeling,vynnycky2010introduction}.
The models in use are enormously varied and employ a wide range of techniques, including
mechanistic mathematical approaches \cite{Gog2020, RSTA-2021-0307}, 
statistical models trained from disease data \cite{Pooley2022, RSTA-2021-0301, RSTA-2021-0302, RSTA-2021-0303}, and 
computational micro-simulation of agents \cite{Panovska-Griffiths2020, Kerr2021, SPOONER2021114461, RSTA-2021-0304, RSTA-2021-0315}
as well as complex model emulators \cite{RSTA-2022-0039}.
Each aims to generate outputs to help understand the past, current, or future course of an epidemic whilst considering context-specific strategies for mitigating spread. 
Mathematical models have limitations and uncertainties, especially when modelling novel infections such as COVID-19. Moreover, different types of models can give different types of insight even when they are describing the same epidemic process \cite{Fyles2021, Danon2021}.
For these reasons it is also important to effectively communicate the underlying model assumptions and the uncertainties in their estimates \cite{Sridharm1567}.

Designing, constructing, and using such models therefore comes with significant challenges.
In responding to an emergency we need to develop good answers to complex questions at speed:
\begin{itemize}
    \setlength{\itemsep}{-1pt}%
\item \textit{Which structures and techniques are most effective and informative?}
\item \textit{How can we be sure there are no bugs or errors?}
\item \textit{How much detail should be included in the model?}
\item \textit{What data should be used as inputs for the model and at what spatial and temporal scales?}
\item \textit{What are the effects of using different data sets and model parameter settings on model outputs? How do these effects vary?}
\item \textit{How can model mechanics and outputs be effectively communicated to diverse audiences?}
\item \textit{How can the models be used to inform and compare policies?}
\item \textit{How can model-informed decisions be effectively and widely communicated?}
\end{itemize}
Efforts to quickly and reliably establish plausible answers to these questions during the COVID-19 pandemic have involved teams of academics working in parallel on complementary models and rapidly developing new collaborations and ways of working to support their research.
Interdisciplinary expertise in, for example, software engineering, spatial analysis, and visualization have been drawn upon as modellers have endeavoured to address the various challenges involved in responding to the pandemic in timely and informative ways.

\subsection{Visualization and Visual Analytics (VIS)}

Visualization emerged as a challenging and enabling application of computer science in the early 1990s.
It offered a paradigm shift in the way that processing power and capabilities for rendering were applied to scientific data \cite{mccormick1987visualization}.
Drawing upon long-standing traditions in
cartography \cite{bertin_semiology_1983, robinson1995elements},
statistics \cite{tukey1977exploratory, cleveland1985elements},
graphic design \cite{neurath1974isotype,arntz2010gerd,jansen2009neurath} and
cognitive science \cite{kahneman1981perceptual},
and leveraging their developing interactions \cite{cleveland1984graphical,tufte1983visual,stuetzle1987plot},
it was perhaps best captured by Mackinlay \cite{mackinlay1986automating}.
This shift sparked cutting edge research in the development of hardware, software, theory and techniques for depicting and interacting with data.
The need for tight integration between graphics and analytical capability, interactive rendering speeds for rapid updates and support for analytic workflows for discovery
has developed out of this initiative \cite{thomas2006visual} as
researchers and designers have
supported data analysts in science, government and industry through visualization and visual analytics.


The diverse forms of knowledge \cite{lee2019broadening} that have resulted include techniques, methodologies and epistemologies that enable VIS to contribute meaningfully and effectively to problems, ranging from highly specialized academic domains to urgent and imminent global challenges.
Rapid and flexible interactions with rich graphical depictions of data
enable us to understand the complexities and nuances of
atmospheric models \cite{treinish1998task, sanyal2010noodles, costa2020interactive},
poetry composition \cite{abdul2013rule, mccurdy2015poemage},
animal ecology \cite{andrienko2013space,slingsby2016exploratory,walker2016timeNotes},
sporting performance \cite{legg2012matchpad,andrienko2017visual,stein2017bring,andrienko2019constructing},
transport systems \cite{andrienko2008spatio,tominski2012stacking,andrienko2012scalable,wood2011visualizing,beecham2014exploring,beecham2014studying
},
evolution \cite{meyer2009mizbee,meyer2010pathline},
cyber attacks \cite{mckenna2015unlocking, angelini2018ROPMate, goodall2018situ}, 
energy consumption \cite{rodgers2011exploring, goodwin2013creative},
healthcare \cite{Elshehaly:2021:TVCG,Gotz2016},
genetics \cite{nielsen2009abyss, ferstay2013variant}, and many other aspects of nature and society including epidemics and epidemiology \cite{guo2007visual,afzal2011visual,maciejewski2011pandemic, Carroll2014,  bryan2015integrating, Dixit:2020:JAMIA, crisan2021gevitrec}.

\subsection{Rapid Assistance in Modelling the Pandemic}

\summary{SCRC responded to RAMP and Vis support was sought / offered}

The \textit{Royal Society} contributed to the interdisciplinary effort to address the COVID-19 pandemic in the UK by convening the \textit{Rapid Assistance in Modelling the Pandemic (RAMP)} initiative \cite{RAMP:2020:web}.
RAMP identified willing volunteers with potential to contribute to the challenges involved in modelling and linked them with candidate epidemiological modelling groups.
% ...
The Scottish COVID-19 Response Consortium (SCRC) formed as part of this RAMP initiative with a focus on
developing a robust understanding of the impacts of different exit strategies from lockdown \cite{scrc2021web}.
Some 150 epidemiologists, software developers, mathematical modellers, data scientists and others came together to do so.
SCRC made a call for visualization volunteers to help with the effort in May 2020.
As a result more than 20 visualization researchers and developers, from 11 universities and 2 companies in the UK, offered to volunteer by providing visualization support for the epidemiological modelling in SCRC.


\summary{Process - what we did!  (Remember - much of the detail is in arXiv / Epidemiology and does not need to be repeated)}

Visualization theory~%
\cite{chen2019ontological} was used 
to structure this activity, specifically to 
establish the broad applicability of visualization in modelling workflows and communicate this persuasively to modelling scientists \cite{chen2021epidemics}.
Activities involving various modelling workflows were supported concurrently (Fig.~\ref{fig:4LOV}) by creating seven groups to work in parallel. They provided either \textit{model specific} or more \textit{generic} support for visualization and associated \textit{analytical capability}.
Four groups of visualization researchers sought close collaboration with specific modelling teams working on particular models.
The intention was to explore opportunities for using graphics to provide rapid feedback on model inputs, outputs and inner workings with the intention of 
addressing the challenges involved in \textit{model development} (\secref{sec:EM}).
Three central visualization support teams provided a complementary more generic focus on:
the rapid development of a \textit{visualization infrastructure} for interactive visualization of the data made available through the SCRC data pipeline \cite{scrcDataPipeline, RSTA-2021-0300};
the visual communication and \textit{dissemination} of the SCRC work to a broader audience, and;
the provision of \textit{analytical capability} to the other groups.

The volunteering took place from June 2020 to January 2021. This initial effort has been consolidated by a follow-on project funded by the UKRI/EPSRC through their \textit{COVID 19 Rapid Response} programme - \textit{RAMP VIS: Making Visual Analytics an Integral Part of the Technological Infrastructure for Combating COVID-19}.
Running until January 2022, this subsequent 12-month project has helped maintain and develop the activity generated through the volunteering, while providing additional resources to enable further capability, research and communication \cite{chen2021epidemics, khan2021propagating}.




\subsection{Supporting Epidemiological Modelling with Visualization Research}
\label{sub:visForEpMod}

\summary{
\textBF{NEW:}
Visualization Knowledge and Applied Visualization Research
Or is this just part of `Approach'?
}

Much of the knowledge generated in visualization research is encapsulated in distinct visual designs.
Munzner \cite{munzner2014visualization} adopts the terminology of scientific visualization pioneers Haber \& McNabb \cite{haber1990visualization} in describing these designs as \emph{idioms}
-- combinations of design decisions that result in particular styles or forms of visual data depiction that are likely to be useful for particular types of user to support particular types of task \cite{miksch2014matter}.
These idioms can be considered as graphical templates that offer (sometimes loosely) defined styles of visual communication.
They include well known graphic types such as:
scatter plots \cite{RSTA-2021-0301} (\textit{Fig. 4});
choropleth maps \cite{RSTA-2021-0304} (\textit{Fig. 2}), \cite{RSTA-2021-0315} (\textit{Fig. 4});
bar charts \cite{RSTA-2021-0307}, (\textit{Fig. 5}), \cite{RSTA-2021-0311} (\textit{Fig. 4});
heat maps \cite{RSTA-2021-0298} (\textit{Fig. 2}), \cite{RSTA-2021-0311} (\textit{Fig. 4}), \cite{RSTA-2022-0039} (\textit{Fig. 3});
time lines \cite{RSTA-2021-0301} (\textit{Fig. 8}), \cite{RSTA-2021-0315} (\textit{Figs. 5,6}), \cite{RSTA-2022-0039} (\textit{Fig. 2}) and
box plots \cite{RSTA-2021-0314} (\textit{Fig. 2}),
as well as many less celebrated and more specific graphical devices \cite{heer2010tour}.
Idioms also involve approaches to generating composite graphics, for example through Small Multiples \cite{RSTA-2021-0301} (\textit{Fig. 7}), \cite{RSTA-2021-0307} (\textit{Fig. 4}), \cite{RSTA-2021-0314} (\textit{Figs. 4,5}), \cite{RSTA-2022-0039} (\textit{Figs. 5, 6})]
Knowledge about how to use visualization effectively in any applied context draws upon these established
visual solutions and their use to address particular problems (task, data set, user) \cite{miksch2014matter}. It is frequently and beneficially shaped in the crucible of applied work.
Some of this knowledge is explicitly captured in a series of complementary efforts to determine and communicate what we know about visualization through established idioms (e.g. \cite{mackinlay2007show, diehl2018visguides, ftDataVocab, ft2021visual}), but much of it remains implicit and is drawn upon, and accumulated, in applied contexts through
\textit{the visualization design process}.
Various models describe this process of understanding tasks, users and data, and applying and developing visualization knowledge in specific beneficial ways.
They detail
the strategies to apply,
the activities in which to engage,
the pitfalls to avoid, 
and the qualities expected in rigorous work \cite{shneiderman2006strategies, lloyd2011human, sedlmair2012design, meyer2012four, wood2014moving, meyer2015nested, mccurdy2016action, hall2019design, meyer2019criteria}.
Munzner \cite{munzner2014visualization} emphasizes the importance of these idioms for knowledge \textit{transfer} between contexts during the design process in noting the crucial role of:
\sayIt{existing idioms as [providing] a springboard for designing new ones}.


Our approaches were framed by these ideas, but, in the spirit of transfer, adapted to the rapid nature of our emergency context. 
Applied visualization research usually involves system design and development that takes place through an ongoing iterative process of redesign, resulting in a usable visualization system.
Typically, complimentary processes occur at different time scales \cite{chen2015may}, including:
\begin{itemize}
    \setlength{\itemsep}{0pt}%
\item \textit{\textbf{initial -- transfer} of known idioms} -- candidate templates that have been shown to be effective in particular combinations of task, user and data \cite{miksch2014matter} are quickly applied to an established problem in which the characteristics of task, user and data are comparable;
\item \textit{\textbf{short-mid term -- ongoing redesign} of visualization prototypes} -- data rich interactive prototypes are rapidly developed to refine these ideas and iteratively redefine problem (supported by task) and solution (idiom informed visualization design) in light of data, user reactions, learning and reflection on task;
\item \textit{\textbf{long-term -- system development} of more stable capability} -- to address identified task(s) in ways that are transferable to other data sets (and tasks) through stable reliable capability, usually in persistent software that applies, refines, combines and gives access to idioms proven to be effective in earlier stages.
\end{itemize}
Much of our initial collaboration involved an iterative process of finding opportunities for beneficially using known graphical approaches, with \textit{Design Study Methodology} \cite{sedlmair2012design} (see Fig.~1) offering good guidance for identifying the kinds of problems where visualization may be
appropriate.
We quickly identified known idioms for transfer from the broad body of visualization knowledge that might be applicable in such cases.
Working at speed, visualization researchers drew upon tools and techniques in which they had expertise to design and deliver solutions that could be injected into the modelling processes and workflows to contribute to discussion and understanding.

The context of the COVID-19 situation involved disruption, but provided urgency and focus.
The digital workplace of lockdown with regular video calls, seminars and other digital knowledge exchange 
enabled, and indeed required
us to \textit{Design by Immersion} \cite{hall2019design}.
This helped visualization researchers to quickly develop an understanding of a series of models (data) being developed in parallel by modelling teams (users), identify questions (tasks) to which known visualization approaches might be usefully applied.
They produced designs that `\textit{springboarded}' existing visualization knowledge and capacity by applying and developing known idioms and appropriate technologies as described above.
The epidemiological modellers learned concurrently about visualization possibilities and practice as visualization was applied to their data and tasks. 
This achieved the kind of
reciprocal immersion, shaping and influence \cite{hall2019design, mccurdy2016action} between epidemiologists, modellers and visualization experts that is so core to effective visualization support and successful applied visualization research \cite{meyer2019criteria}.
We considered and reassessed needs and capability through this process of iterative transfer and redesign
as knowledge of the models and possible disease progression was established at pace.

\subsection{Challenges, Solutions, Reflection \& Recommendations}

\summary{We are trying to introduce VIS here and learn about it by reflecting on the RAMPVIS collaboration}

\summary{
Results - what have we actually produced : software, knowledge, connections, funding. How best to summarise this? 
}

This activity, and our subsequent consolidation through RAMP VIS, has resulted in a central visualization server that offers:
hundreds of
plots and composite dashboards depicting data on the core pandemic indicators \cite{Khan:2022:TVCG,khan2022rapid} (Fig.~\ref{fig:propagation});
analytical agents that automatically transform raw data to be visualized by the central system;
a collection of analytical routines and algorithms that offer generic analytical capability for exploring time-series based data;
a series of static and interactive visualization prototypes to support the four modelling teams.
Work on the modelling support prototypes has resulted in, for instance:
new ways of representing and interacting with data for contact tracing and assessing model inputs and outputs (Fig.~\ref{fig:teaser});
improved epidemiological models and understanding of them (Fig.~\ref{fig:GGM});
new connections between research groups and researchers;
new attitudes to the use of visualization in epidemiological modelling and shared knowledge about how this can take place, and;
funding to support ongoing work.
%
%
In line with SCRC protocols we used open source solutions to ensure transparency, invite scrutiny and prepare for reuse \cite{meyer2019criteria}.
A series of academic articles, both published and in preparation, in the visualization and epidemiological modelling domains, report on specific results that combine technology, design and data successfully in new and revealing ways, e.g. \cite{chen2021epidemics, Khan:2022:TVCG,khan2022rapid}.

% ...

\summary{Intention}


This paper does not focus in detail on the specific developments that took place. While we document some of these in ways that make them open for potential transfer in the
supplementary materials
that are a key contribution in this publication \cite{dykes_zenodo5717367}, our main intention is to report on the work more broadly and learn from the collective experience of the SCRC volunteering and our subsequent efforts to consolidate.
We aim to capture knowledge that emerges by stepping back and sampling from the mass of parallel activity that has occurred to provide a project-wide view.
Doing so is intended to develop knowledge about visualization in epidemiological modelling during emergency response 
that may also be used in other domains and contexts -- perhaps other epidemiological emergencies, perhaps other kinds of emergency, perhaps other forms of applied visualization.
This kind of ambitious and challenging (probably unique and inherently partial) meta-analysis is enables us to shine a light on what visualization can offer as a discipline and on how visualization research and researchers can be used in epidemiology, and by inference elsewhere, to support and enhance scientific activity with informative interactive analytical graphics.
% ...
The intention is thereby twofold -- to use some of the rich and diverse experiences of the RAMPVIS collaboration:
\begin{itemize}
  \setlength\itemsep{0pt}
\item to \textit{present visualization knowledge and practice to a broad scientific audience} through example, and
\item to \textit{contribute new knowledge to the visualization community} through experience.
\end{itemize}
These are not unrelated aims, and we address them by
documenting \textit{challenges} encountered and \textit{solutions} developed during the collaborative activity.
We do so in the context of a series of existing \textit{knowledge constructs} selected from the VIS literature.
We use these to \textit{reflect} collectively upon the activity and make \textit{recommendations} to support and develop ongoing collaboration and research.
This enables us to assess whether, how and where visualization can be effective,
identify open problems emerging from the efforts, provide guidance for future collaborative activity and make recommendations for developing the visualization community and visualization capacity in ways that can be beneficial in epidemiology and beyond.
Our intention is to use this paper and its
supplementary materials
to showcase VIS and its role in observation, analysis, model development and dissemination in epidemiological modelling.
In doing so, we demonstrate directly the value of the kind of inter-disciplinary collaboration supported by RAMP
and by inference, the potential for use of visualization approaches, knowledge and capacity in other domains. 

\section{Approach} 
\label{sec:approach}

\summary{
We explain what we have done with this experience to try to make sense of it and write it up.
This is the methodology section.
How did we develop what we have here - and why should people take it seriously. This is all about establishing internal validity.
How did we come up with the findings that follow?
How are we able to claim some confidence in them?
}

Given the rapid, distributed, dynamic, unpredictable and sometimes chaotic nature of the engagement, in which researchers were volunteering their time while interacting with new colleagues and new domains, whilst experiencing the pandemic themselves, our knowledge claims are not underpinned by a deeply planned data collection methodology.
However, we have taken time and steps to develop and coordinate reflection upon the individual and collective experiences - some of which were not shared until we began this very process.
Doing so enables us to make tentative suggestions about visualization, visualization design processes and how visualization might be used more fully in epidemiological modelling and perhaps other domains in the future. This speculation is based upon discussion among the researchers involved, with cross-pollination of ideas and reactions, both to the visceral experiences of the summer of 2020 and the subsequent work in supporting epidemiological modelling through visualization.

\subsection{Selection and Reflection on Constructs}
\label{sub:constructReflectionOn}

We used a structured process to document and reflect on our engagement.
The methods are in line with 
\textit{Design by Immersion}  activities \activity{S}{2}{document} and \activity{P}{5}{critique} \cite{hall2019design},
which resonate with key concepts used in other visualization design process models such as the
\textit{guided emergence} principle of \textit{Action Design Research} \cite{mccurdy2016action}
and the \textit{record} and \textit{reflect} recommendations for interpretivist design study \cite{rogers2020insights}.

In an initial reflective activity, visualization volunteers were asked to identify and select specific examples of established visualization knowledge that they used in initial transfer as the springboard for supporting epidemiological modelling.
We asked the volunteers to select idioms and other knowledge constructs that they applied, and use these as a basis for documenting and reflecting upon their experiences and establishing their learning.
We define a visualization knowledge construct, henceforth `\textit{construct}', as: 
\begin{itemize}
    \setlength{\itemsep}{0pt}%
\item \sayIt{something that is explicitly known and understood in the visualization body of knowledge
that was transferred to address a problem in the current context of visualization support for epidemiological modelling.}
\end{itemize}

Given the importance of collaboration and communication to the endeavour, and previous positive experience \cite{wood2018design, beecham2020design} we adopted the \textit{computational notebook} paradigm for documentation of these constructs and reflection upon them.
Notebooks encourage explicit reasoning around
design choices with explanatory text surrounding embedded code that implements and renders graphics.
Based on the ideas of Literate Programming \cite{knuth1984literate} and the computational essay \cite{wolfram2017computational}, they support activities and principles that are core to effective applied visualization design.
Examples include
\activity{C}{2}{common knowledge},
\activity{C}{3}{peer-to-peer communication} and
\activity{C}{4}{translate concepts}
\cite{hall2019design}, and 
\activity{P}{3}{reciprocal shaping} and \activity{P}{5}{authentic and concurrent evaluation} \cite{mccurdy2016action}.
Use of these activities and adherence to these principles offers a form of internal validation \cite{meyer2019criteria} in supporting information exchange and discussion between collaborators.

The result of this collective reflective activity is \numNotebooks commonly structured but independently authored digital notebooks.
They are listed in Table \ref{tab:suppMaterials} and available for detail and scrutiny through the live links provided in this document.
They focus on specific constructs selected by volunteers as a frame for reporting and reflecting on their contributions and experiences.
Each explains the context in which the construct was applied and the problem it was intended to solve through a `\textit{User Story}', before documenting what is known about it, why it was considered a candidate for transfer,
how it was used and how it performed.
Potential for further transfer is supported in many cases through links to illustrative examples of the construct in use beyond the current context, along with full sources that describe the original research.
Critically reflective thinking \cite{brookfield1998critically, meyer2018reflection} on the transfer and use of the construct to support the epidemiological modelling is then recorded as researchers document their experience and assess what was learned as a result.
% ...
We use \textit{Observable} \cite{observable} (formerly \textit{d3.express}) as the computational notebook platform to benefit from its emphasis on reactive literate visualization \cite{wood2018design}, which supports visualization-rich discourse.
This approach and technology encourages reflective reasoning in something approximating a \textit{reflective schema} \cite{meyer2018reflection} and makes it explicit and available both for analysis as part of the design discourse and for wider scrutiny as we assess and synthesize experiences to develop broader claims.

\begin{figure*}[htb]
  \centering
  \vspace{4px}
  \includegraphics[width=.80\textwidth]{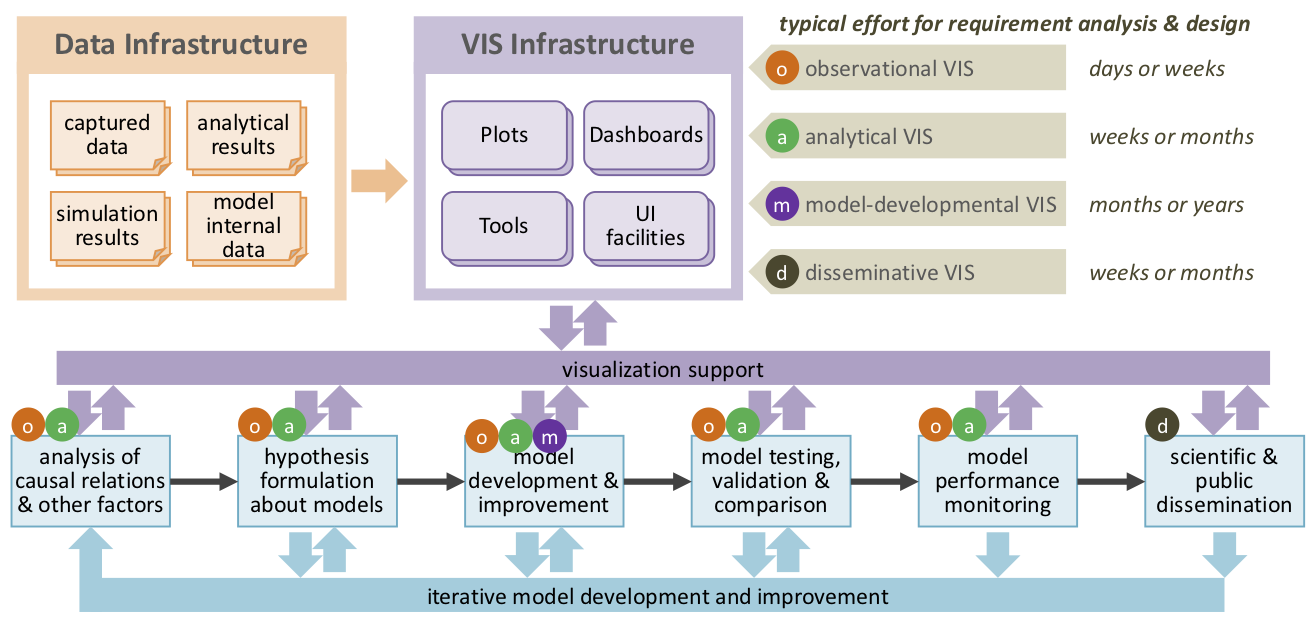}
  \caption{
  Project architecture expressed using the \textit{Four Levels of Visualization} \cite{https://observablehq.com/d/a9aaed2f31718620} model. Structuring and planning our engagement in light of the model \cite{chen2015may} helped us coordinate the SCRC volunteering effectively, efficiently and flexibly.
  By identifying needs for disseminative, observational, analytical \& model developmental visualization we were able to identify six activities,  and plan for the different amounts of effort needed for each. This enabled us to deploy the available VIS volunteers based on their expertise appropriate for each level of tasks and provide flexible visualization support across SCRC
  \protect\obs{https://observablehq.com/d/a9aaed2f31718620,729}.
  We used the model to develop an iterative approach to establishing opportunities, supporting and developing needs, prototyping solutions and reflecting on action that underpins this paper.
  }
  \label{fig:4LOV}
\end{figure*}

The diverse forms of contribution made by volunteers in the various roles in the project and the variety of efforts to which they were dedicated (see Fig.~\ref{fig:4LOV}) enables us to consider a range of forms of visualization knowledge construct.
Some are idioms, visual design templates in the Munzner \cite{munzner2014visualization} and Haber \& McNabb \cite{haber1990visualization} sense, where the focus is predominantly on transformations from data to graphics.
But we capture a wider range of visualization knowledge types, such as:
enabling technologies \obs{https://observablehq.com/d/071ee158d5418d96},
theoretical models \obs{https://observablehq.com/d/54c8641168c013ea},
design methodologies \obs{https://observablehq.com/d/2e98f8d7f3cf5c08} \obs{https://observablehq.com/@jsndyks/rampvis-idiom-design-by-immersion},
narrative devices \obs{https://observablehq.com/@ritsosp/rampvis-idioms-narrative-design-patterns}
analytical approaches \obs{https://observablehq.com/d/78b20aa4152547e2} and
system architecture configurations \obs{https://observablehq.com/d/596df309c41cca50} in this rich data source.



\subsection{Synthesis and Reflection Across Constructs}
\label{sub:constructReflectionAcross}

In a second stage of critically reflective activity we consider, compare and contrast these post-event reflections `\textit{on action}' \cite{schon1987educating}.
This enables us to suggest key themes that emerge
through an inductive process
that aligns with the relativist approach to knowledge generation taken here \cite{smith2017developing, meyer2018reflection, meyer2019criteria}.
The stages of analysis were not entirely separate.
Each notebook was developed iteratively, with the individual experiences prompted and informed by the higher level concepts as they emerged as part of the process of collective reflection.
Feedback to notebook authors (that we retain as a record in the
supplementary materials) enabled us to prompt for ideas, question, refine and improve descriptions, calibrate where possible, communicate effectively and relate experiences to emerging cross-cutting themes as we developed our evidence base, reflected upon its characteristics and synthesized the diverse views and experiences.
This reflection on the `data' captured in our notebooks is the basis for a series of cross-project findings.

Sedlmair\etal~regard reflection as being \sayIt{where research emerges from engineering} and implementing this reflective process enables us to document the transfer of some of what is known in visualization to the domain of epidemiological modelling and test and develop this knowledge in the crucible of the pandemic response. Given this approach, we offer no claims of absolute and indefatigable truth, but are careful to relate themes and claimed findings to specific evidence and particularly value cases where multiple experiences converge in our cross-construct meta-analysis.


\begin{figure*}[htb]
  \centering
  \vspace{4px}
  \includegraphics[width=.99\textwidth]{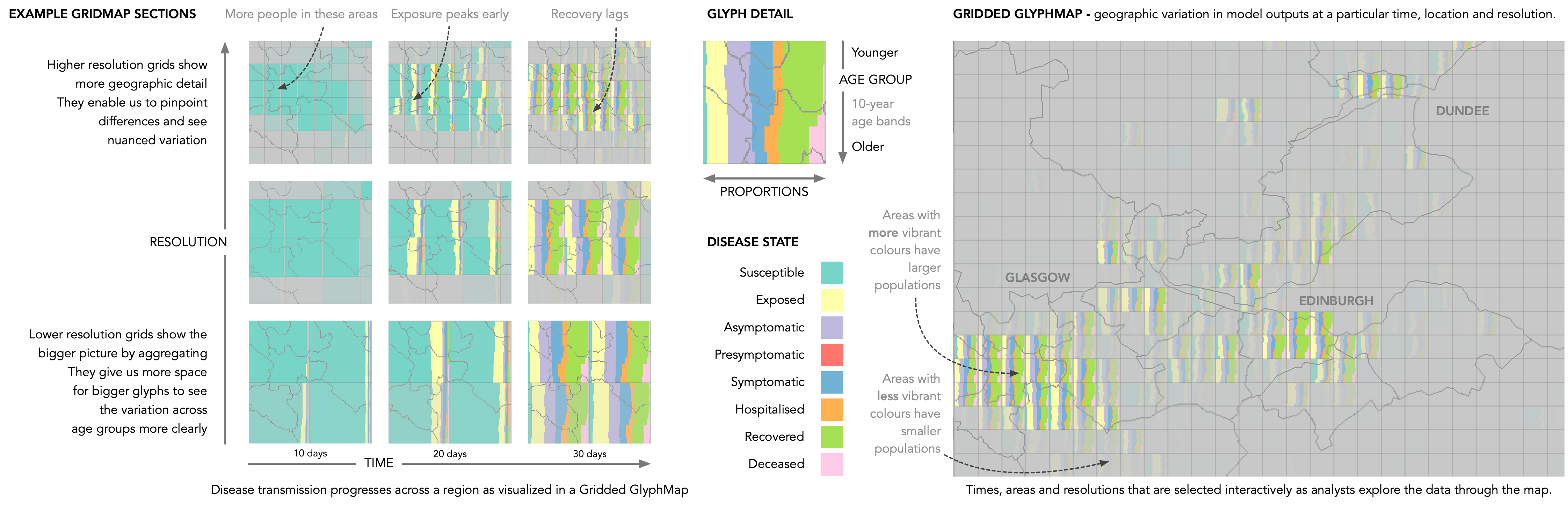}
  \caption{
  Nine excerpts (\textit{left}) of an interactive exploration of a modelled disease outbreak visualized using our \textit{GriddedGlyphMap} prototype \cite{https://observablehq.com/@aidans/rampvis-idiom-gridded-glyphmaps}.
  Cells represent areas in Scotland, with areas of higher populations being shown in more vibrant colours. Each glyph shows proportions of population in particular disease states (colours, horizontal proportions) in 10-year age bands (vertical rows), revealing the spatial and age-based characteristics of a modelled disease outbreak.
  The large image (\textit{right}) shows a wider spatial overview of a single time-point at a particular scale.
  Modellers interacted with the output data to reveal patterns that resulted in changes to the model code and deeper understanding of the effects of changes to the model as knowledge of the disease progressed.
  }
  \label{fig:GGM}
\end{figure*}

\section{Findings -- Results and Claims}

\summary{
We make some claims based upon the experience and our reporting and analysis of it. These fall out of the reflection and are based on our synthesis of that. Initial analysis and efforts at synthesis suggests that proposed claims / findings are: \\
\textbf{VIS can work well} (introduces some idiom examples, adds some types of success);
\textbf{But not Always} (learning about how and when to use VIS);
\textbf{Experimental Approach} is a useful framing to achieve the above);
\textbf{Importance of Community Knowledge};
\textbf{Inadequacy of Current Process Models};
\textbf{Importance of Applied Research};
}

The diverse array of constructs, listed in Table~\ref{tab:suppMaterials} and described fully in our supplementary materials \cite{dykes_zenodo5717367}, gives us a broad and informative set of experiences of using visualization methods and approaches in
emergency response.
This is our evidence base.
We provide direct access to it through \textit{hyperlinks} to the cited notebooks themselves (e.g. \obs{https://observablehq.com/@jwolondon/rampvis_idiom_odmap}) or \textit{specific cells} within them that contain explanations, examples, graphics and reflections (e.g. \obs{https://observablehq.com/@jwolondon/rampvis_idiom_odmap,1912}).
It is something of a \textit{trrrace} \cite{rogers2020insights} through which we \textit{r}ecord, \textit{r}eport and \textit{r}eflect on our activity.


\label{tab:suppMaterials}
\begin{center}
\begin{table}
\renewcommand{\arraystretch}{1.15}
\rowcolors{1}{gray!2}{gray!8}
\begin{tabular}{ | r | c | l | } 
 \hline



{\footnotesize\textbf{\href{https://observablehq.com/d/071ee158d5418d96}{D3 Prototyping}}} & \obs{https://observablehq.com/d/071ee158d5418d96} & {\fontfamily{cmss}\selectfont\small An expressive JavaScript library for VIS}\\
{\footnotesize\textbf{\href{https://observablehq.com/d/78b20aa4152547e2}{New Data Spaces}}} & \obs{https://observablehq.com/d/78b20aa4152547e2} & {\fontfamily{cmss}\selectfont\small Explore data in transformed spaces }\\
{\footnotesize\textbf{\href{https://observablehq.com/d/2e98f8d7f3cf5c08}{VIS Design Study}}} & \obs{https://observablehq.com/d/2e98f8d7f3cf5c08} & {\fontfamily{cmss}\selectfont\small Design{\tinySpace}methods for emergency{\tinySpace}response}\\
{\footnotesize\textbf{\href{https://observablehq.com/@aidans/rampvis-idiom-gridded-glyphmaps}{Gridded GlyphMap}}} & \obs{https://observablehq.com/@aidans/rampvis-idiom-gridded-glyphmaps} & {\fontfamily{cmss}\selectfont\small Animated{\tinySpace}maps of model{\tinySpace}outputs by age}\\
{\footnotesize\textbf{\href{https://observablehq.com/@ritsosp/rampvis-idioms-pictograms}{Pictograms}}} & \obs{https://observablehq.com/@ritsosp/rampvis-idioms-pictograms} & {\fontfamily{cmss}\selectfont\small Pictorial symbols for representing data}\\
{\footnotesize\textbf{\href{https://observablehq.com/@kaimdx/rampvis-idiom-gmap}{GMap}}} & \obs{https://observablehq.com/@kaimdx/rampvis-idiom-gmap} & {\fontfamily{cmss}\selectfont\small Visualizing network clusters as regions}\\
{\footnotesize\textbf{\href{https://observablehq.com/@ritsosp/rampvis-idioms-narrative-design-patterns}{NAPA Cards}}} & \obs{https://observablehq.com/@ritsosp/rampvis-idioms-narrative-design-patterns} & {\fontfamily{cmss}\selectfont\small Narrative patterns for VIS storytelling}\\
{\footnotesize\textbf{\href{https://observablehq.com/d/7d339207ef90c483}{Linked PPLots}}} & \obs{https://observablehq.com/d/7d339207ef90c483} & {\fontfamily{cmss}\selectfont\small Explore{\tinySpace}relationships{\tinySpace}between{\tinySpace}parameters}\\
{\footnotesize\textbf{\href{https://observablehq.com/d/43927395f6cb890c}{Alternative Views}}} & \obs{https://observablehq.com/d/43927395f6cb890c} & {\fontfamily{cmss}\selectfont\small Many{\tinySpace}lenses to explore dynamic{\tinySpace}networks }\\
{\footnotesize\textbf{\href{https://observablehq.com/d/4aebb875cacaef3a}{Visual Faceting }}} & \obs{https://observablehq.com/d/4aebb875cacaef3a} & {\fontfamily{cmss}\selectfont\small Cut, sort, align and compare time-series}\\
{\footnotesize\textbf{\href{https://observablehq.com/@lborohfang/rampvis-idiom-integrated-algorithmic-tools-for-visual-ana }{VA for Modelling}}} & \obs{https://observablehq.com/@lborohfang/rampvis-idiom-integrated-algorithmic-tools-for-visual-ana } & {\fontfamily{cmss}\selectfont\small Introduce VA to modelling workflows}\\
{\footnotesize\textbf{\href{https://observablehq.com/d/87a416cd4468fff0}{Propagating{\tinySpace}Designs}}} & \obs{https://observablehq.com/d/87a416cd4468fff0} & {\fontfamily{cmss}\selectfont\small Automatic generation of visualizations}\\
{\footnotesize\textbf{\href{https://observablehq.com/@jwolondon/rampvis_idiom_directional_flow_curves}{Flow Curves}}} & \obs{https://observablehq.com/@jwolondon/rampvis_idiom_directional_flow_curves} & {\fontfamily{cmss}\selectfont\small Map directional flows between regions}\\
{\footnotesize\textbf{\href{https://observablehq.com/@henryqw/multiple-linked-views}{Linked Views}}} & \obs{https://observablehq.com/@henryqw/multiple-linked-views} & {\fontfamily{cmss}\selectfont\small Multiple{\tinySpace}view{\tinySpace}parameter{\tinySpace}space{\tinySpace}exploration}\\
{\footnotesize\textbf{\href{https://observablehq.com/d/596df309c41cca50}{Ontology for VIS}}} & \obs{https://observablehq.com/d/596df309c41cca50} & {\fontfamily{cmss}\selectfont\small Knowledge center of a{\tinySpace}VIS{\tinySpace}infrastructure}\\
{\footnotesize\textbf{\href{https://observablehq.com/d/087e459840b2b478}{Workflow Analysis}}} & \obs{https://observablehq.com/d/087e459840b2b478} & {\fontfamily{cmss}\selectfont\small Analyse |cause|remedy|side-effect}\\
{\footnotesize\textbf{\href{https://observablehq.com/d/c3a628d8bd4fe8fe}{Parallel Plots}}} & \obs{https://observablehq.com/d/c3a628d8bd4fe8fe} & {\fontfamily{cmss}\selectfont\small PCP for parameter summary{\tinySpace}\&{\tinySpace}selection}\\
{\footnotesize\textbf{\href{https://observablehq.com/d/ea78b530cac2ee60}{Error Encodings}}} & \obs{https://observablehq.com/d/ea78b530cac2ee60} & {\fontfamily{cmss}\selectfont\small Error vs uncertainty in parameter space}\\
{\footnotesize\textbf{\href{https://observablehq.com/@jcrbrts/rampvis-idioms-fds}{Five Design Sheets}}} & \obs{https://observablehq.com/@jcrbrts/rampvis-idioms-fds} & {\fontfamily{cmss}\selectfont\small Structured sketching approach{\tinySpace}to{\tinySpace}design}\\
{\footnotesize\textbf{\href{https://observablehq.com/d/a9aaed2f31718620}{Four Levels of VIS}}} & \obs{https://observablehq.com/d/a9aaed2f31718620} & {\fontfamily{cmss}\selectfont\small Categorize VIS task complexity \& effort}\\
{\footnotesize\textbf{\href{https://observablehq.com/@jwolondon/rampvis_idiom_odmap}{OD Maps}}} & \obs{https://observablehq.com/@jwolondon/rampvis_idiom_odmap} & {\fontfamily{cmss}\selectfont\small Show geography in interaction matrices}\\
{\footnotesize\textbf{\href{https://observablehq.com/d/880d234e47a1ed24}{Scalable{\tinySpace}Pixel{\tinySpace}Views}}} & \obs{https://observablehq.com/d/880d234e47a1ed24} & {\fontfamily{cmss}\selectfont\small Show variation of each parameter value}\\
{\footnotesize\textbf{\href{https://observablehq.com/d/e635bbb87b89b463}{Dynamic Data Vis}}} & \obs{https://observablehq.com/d/e635bbb87b89b463} & {\fontfamily{cmss}\selectfont\small Automatic{\tinySpace}agents update{\tinySpace}VIS{\tinySpace}systems }\\
{\footnotesize\textbf{\href{https://observablehq.com/d/54c8641168c013ea}{Infrastructure}}} & \obs{https://observablehq.com/d/54c8641168c013ea} & {\fontfamily{cmss}\selectfont\small Ensure readiness in emergency{\tinySpace}response}\\
{\footnotesize\textbf{\href{https://observablehq.com/d/e7dc9a1ffd763d73}{Deployment}}} & \obs{https://observablehq.com/d/e7dc9a1ffd763d73} & {\fontfamily{cmss}\selectfont\small Develop reliable visualization systems}\\

{\footnotesize\textbf{\href{https://observablehq.com/d/bb1d4a08de896005}{VIS Guidelines}}} & \obs{https://observablehq.com/d/bb1d4a08de896005} & {\fontfamily{cmss}\selectfont\small Extract, record, transfer VIS knowledge }\\
{\footnotesize\textbf{\href{https://observablehq.com/d/ac04bb76a1ec0dd4}{VIS Theory}}} & \obs{https://observablehq.com/d/ac04bb76a1ec0dd4} & {\fontfamily{cmss}\selectfont\small Knowledge-based{\tinySpace}VIS{\tinySpace}requirement{\tinySpace}analysis}\\
{\footnotesize\textbf{\href{https://observablehq.com/d/efe6c1c90c625bec}{Volunteer VIS}}} & \obs{https://observablehq.com/d/efe6c1c90c625bec} & {\fontfamily{cmss}\selectfont\small Provide VIS support in emergency }\\

{\footnotesize\textbf{\href{https://observablehq.com/@jsndyks/rampvis-idiom-design-by-immersion}{Design{\tinySpace}by{\tinySpace}Immersion}}} & \obs{https://observablehq.com/@jsndyks/rampvis-idiom-design-by-immersion} & {\fontfamily{cmss}\selectfont\small Activities for Trans-disciplinary VIS}\\

\hline
\end{tabular}
\vspace{4px}
\caption{\label{tab:table-name}RAMP VIS Knowledge Constructs: selected examples of visualization knowledge used to reflect on visualization support for epidemiological modelling. 
Interactive notebooks describing each are available as online supplementary materials \cite{dykes_zenodo5717367}.\\
\label{tab:suppMaterials}
}
\end{table}
\end{center}

This abundant evidence base of diverse 
experiences addresses the overarching tacit hypothesis with which we approached the work:\\
\begin{itemize}
    \item \sayIt{visualization can be effective in epidemiological modelling}.
\end{itemize}
It leads to some specific findings about when and where this may be the case and how it might be achieved.
Relating experiences across activities through the structured reflection and synthesis described in \secref{sec:approach} allows us to develop some more general findings and claims and speculate a little about possible transfer to other related problems, domains, data sets and contexts.
These enable us to:
\begin{itemize}
    \setlength{\itemsep}{0pt}%
  \item add to the growing body of evidence supporting the \textit{utility and applicability} of visualization in the analysis of complex data sets across domains, and understand \textit{how this can be achieved};
  \item identify \textit{open problems} for consideration by the visualization research community as it engages in a wide set of problems and domains;
  \item suggest approaches for an effective \textit{process of engagement, analysis and design} when domain specialists are working with visualization researchers and designers to problem solve with an emphasis on emergency response;
  \item make \textit{recommendations} for the VIS and wider scientific and academic communities, in terms of actions that can be taken and priorities that can be established, for more widespread and informed use of visualization in scientific discovery and emergency support in the future.
\end{itemize}




We begin with six broad claims and evidence to support them.

\subsection{VISUALIZATION CAN BE EFFECTIVE}
\label{sec:success}

Our experiences, efforts and reflection show that visualization can contribute effectively at various levels in the epidemiological context. This is the case not only for communicating results, but to address the questions outlined at the outset (\secref{sec:EM}) by observing, analysing and modelling in line with the \textit{Four Levels of Visualization} (disseminative, observational, analytical \& model developmental) \cite{chen2015may, https://observablehq.com/d/a9aaed2f31718620} used to frame and coordinate the engagement (see Fig.~\ref{fig:4LOV}).
For example, by respectively:

\begin{itemize}
    \setlength{\itemsep}{0pt}%
\item presenting information in ways likely to engage and inform broadly \obs{https://observablehq.com/@ritsosp/rampvis-idioms-narrative-design-patterns,766}, \cite{https://observablehq.com/@jcrbrts/rampvis-idioms-fds}; 
\item applying numeric and graphical transformations to reveal structure \obs{https://observablehq.com/@jwolondon/rampvis_idiom_odmap,2072}, \obs{https://observablehq.com/d/78b20aa4152547e2,684} - see Fig.~\ref{fig:teaser}; 
\item showing relationships between time series and places at various scales \obs{https://observablehq.com/@aidans/rampvis-idiom-gridded-glyphmaps,788},
\obs{https://observablehq.com/@jwolondon/rampvis_idiom_directional_flow_curves,938},
\obs{https://observablehq.com/d/4aebb875cacaef3a,669} - see Fig.~\ref{fig:GGM}; 
\item comparing large numbers of model inputs and outputs at multiple scales \cite{https://observablehq.com/d/ea78b530cac2ee60, https://observablehq.com/d/880d234e47a1ed24}, \obs{https://observablehq.com/@lborohfang/rampvis-idiom-integrated-algorithmic-tools-for-visual-ana,170} 
\end{itemize}

Existing visualization idioms provide plenty of potential for rapid deployment and effective application to epidemiology.
This holds in the case of considerable complexity \cite{https://observablehq.com/d/4aebb875cacaef3a, https://observablehq.com/d/7d339207ef90c483,https://observablehq.com/d/880d234e47a1ed24}, dynamism \cite{https://observablehq.com/d/e635bbb87b89b463,https://observablehq.com/d/43927395f6cb890c, https://observablehq.com/@aidans/rampvis-idiom-gridded-glyphmaps} and uncertainty \cite{https://observablehq.com/d/ea78b530cac2ee60}.
Much of what was achieved was novel in terms of the designs developed, algorithms produced and the epidemiological context to which the ideas were applied - such as the use of interactive linked views to represent uncertainty and error:
\sayIt{a first under the epidemiological setting} as captured in \obs{https://observablehq.com/d/ea78b530cac2ee60,427}

We characterise these successes by presenting six ways in which our documented reflection provides evidence to support the view that visualization can be effective.

\subsubsection{Achieving Important Insights}

Our collaborations helped us understand the models and the disease, and also the capabilities, limitations and scope of some of the visualization designs and approaches. These insights were often mutually shaped \cite{mccurdy2016action} and interdependent.

Domain insights -- knowledge \textit{about epidemiology and models} -- involved, among other things:
\begin{itemize}
  \setlength\itemsep{0pt}
  \item refining different models, \obs{https://observablehq.com/d/43927395f6cb890c,592}, in ways that had not been picked up in standard static graphics \obs{https://observablehq.com/@aidans/rampvis-idiom-gridded-glyphmaps,788} (see Fig.~\ref{fig:teaser}).;
  \item understanding the effects of varying 
  model inputs \obs{https://observablehq.com/@jwolondon/rampvis_idiom_odmap,310};
  \item revealing the complex and subtle geographic characteristics of the modelling \obs{https://observablehq.com/@aidans/rampvis-idiom-gridded-glyphmaps,788} (see Fig.~\ref{fig:GGM}).
\end{itemize}

Equally, design and process insights -- knowledge \textit{relevant to visualization} -- that may be reusable were achieved, such as:
\begin{itemize}
  \setlength\itemsep{0pt}
  \item  the population weighted fading of glyphs used to make more populated areas more salient and effectively \sayIt{focus the eye on areas where population numbers are higher for relative changes} \obs{https://observablehq.com/@aidans/rampvis-idiom-gridded-glyphmaps,788} (as explained in Fig.~\ref{fig:GGM});
  \item a new semi-automated approach to mapping data sets with particular characteristics to plausible visualization designs for dissemination \obs{https://observablehq.com/d/596df309c41cca50,674} (see Fig.~\ref{fig:propagation});
  \item a method for validating such recommendations \obs{https://observablehq.com/d/87a416cd4468fff0,699};
  \item confirmation of the effective use of narrative patterns \obs{https://observablehq.com/@ritsosp/rampvis-idioms-narrative-design-patterns};
  \item an automated process for daily data updates to visualization dashboards for dynamic data dissemination \obs{https://observablehq.com/d/e635bbb87b89b463}.
\end{itemize}
We say more about the \textit{visualization design process} in section~\ref{sub:visprocess} (with a particular focus in~\ref{sub:analysisisdesign}), but can claim with some confidence that:

\effective{Visualization resulted in insights about epidemiology and the design and use of data graphics}

\subsubsection{Capturing and Communicating Complexity}

Much of our reported activity demonstrates scope for showing highly dimensional data in meaningful ways that enable patterns to be detected and models to be understood \obs{https://observablehq.com/d/7d339207ef90c483,867},
\obs{https://observablehq.com/@henryqw/multiple-linked-views,427}.
We used existing approaches to design and develop workable representations of, for example, model outputs that varied 16 parameters with 160 different parameter configurations that were generated through 1,000 stochastic runs for 8 age groups.
This resulted in 20,480,000 time series of 200 days each \obs{https://observablehq.com/d/c3a628d8bd4fe8fe,339},  \obs{https://observablehq.com/d/880d234e47a1ed24,346}.
We applied and refined idioms to reveal structure in complex spatial relationships to help understand and select interaction networks \obs{https://observablehq.com/@jwolondon/rampvis_idiom_directional_flow_curves,1375}.
Our graphics present informative (often interactive - see \secref{sub:rapidlyrefine}) visual overviews and allow analysts
to interpret input data and parameter configurations - a considerable challenge \cite{RSTA-2021-0298} - and relate them to output patterns and variations.
As such:

\effective{Complexity is captured effectively with visualization}

\subsubsection{Detecting Variation, Difference and Change}
\label{sub:variation}

Graphics that show \textit{variations} are particularly useful.
These may be in scale, over time, in outputs based upon stochastic modelling or in input parameter configurations. 
Applying idioms to help modellers detect and consider variation and change by rapidly comparing and assessing informative depictions of differences in data sets was effective.
In \obs{https://observablehq.com/@aidans/rampvis-idiom-gridded-glyphmaps,788} we hear that:
\sayIt{The ability to investigate the differences between different model outputs is critical, and has allowed us to identify problems with the model code itself, and to verify the impact of changes to our understanding of the disease on model outputs.} 

We produced coherent graphics designed for \textit{comparison} \cite{gleicher2011visual} broadly, including examples that successfully explore:
\begin{itemize}
  \setlength\itemsep{0pt}
\item individual parameter configurations and clusters of similar configurations \obs{https://observablehq.com/d/880d234e47a1ed24,427};
\item model outcomes and their variation through uncertainty visualization \obs{https://observablehq.com/d/ea78b530cac2ee60,400};
\item model outputs under particular conditions -- such as  location or age group \obs{https://observablehq.com/@aidans/rampvis-idiom-gridded-glyphmaps,771}; 
\item the effects of the scale at which data were aggregated, which in one case \sayIt{revealed unexpected dependency on the scale of spatial aggregation on the model behaviour}
\obs{https://observablehq.com/@jwolondon/rampvis_idiom_directional_flow_curves,1447}.
\end{itemize}

These and multiple other examples and experiences support a strong claim that: 
\effective{Visualization is highly effective for comparison}

\subsubsection{Delivering Detail}
\label{sub:detail}

Graphics that inform decision-making and act as the basis for reporting often involve reducing data sets to summary statistics for simplicity.
As we have seen, visualization offers richer, nuanced, informative  alternatives that remain comprehensible while delivering detail \obs{https://observablehq.com/@jwolondon/rampvis_idiom_odmap,276}.
%
%
We find compelling examples documented in the notebooks in which visualizations `unpack' such indicators or metrics in informative ways to support the decision making processes at different stages of the modelling workflow.

For instance, in cases where the progression of the pandemic within different simulation runs are compared using their respective R values \cite{anderson1992infectious,nishiura2009effective}.
Here, disaggregated visualizations of infection events led to more fine-grained observations such as that \sayIt{the number of random infections were too high which led to several small infection chains} \obs{https://observablehq.com/d/43927395f6cb890c,592}.
This not only leads to an improved understanding of the epidemiological model and complex transmission patterns, but also makes 
\sayIt{comparisons between simulation runs which capture different policies} a much more informed activity \obs{https://observablehq.com/d/43927395f6cb890c,592}. Figure~\ref{fig:teaser} provides an example.
Similarly, ensembles of simulations are more closely inspected and fully evaluated through visualizations that depict the input-output relationships (\obs{https://observablehq.com/d/ea78b530cac2ee60,427}) rather than relying on aggregate statistics of indicators such as the total number of hospitalisations that a model is forecasting within a given period.
In cases involving the comparative analysis of time-series, visualizations served as effective tools to determine the \textit{portions} of time-series that contribute most fully to summary measures such as autocorrelation \obs{https://observablehq.com/d/78b20aa4152547e2,676}.
Heat map visualizations of the distribution of raw time-series values for the members of a time-series cluster also deliver detail effectively.
The graphics reveal patterns that allow the clusters to be described  through observations such as \sayIt{Cluster-1, the largest of the groups is the most typical 2-peak UTLAs with a much more pronounced 2nd peak} \obs{https://observablehq.com/d/4aebb875cacaef3a,720}.
In these various examples:

\effective{Detailed data are made interpretable with visualization}

\subsubsection{Transforming Data}
\label{sub:transform}

Deriving new data spaces through informative analytical transformations of data and exploring these graphically \sayIt{reveals relations and structures not easily visible within the raw data} \obs{https://observablehq.com/d/78b20aa4152547e2,684}.
The \textit{Critical Creative Culture} that we established (\secref{sub:creativeculture}) resulted in various informative graphical transformations. The use of networks to show the similarities and differences between multiple time-series is an example \obs{https://observablehq.com/d/78b20aa4152547e2,676}.
Similarly, our visualization of the principal components of a multivariate parameter space enabled informative \sayIt{identification of which parameter settings generate more variations in its simulation and which settings generate more similar outputs} \obs{https://observablehq.com/@lborohfang/rampvis-idiom-integrated-algorithmic-tools-for-visual-ana,427}.
Network layouts that use space to show structure proved effective \obs{https://observablehq.com/@kaimdx/rampvis-idiom-gmap,814}.

Some transformations are less abstract, such as \textit{OD maps} \cite{wood2010visualization}.
These flow maps address the overplotting associated with spatially concentrated geographic data by stretching and warping geographic space.
We used this transforming idiom to 
reveal differences in flows of people between locations used as an input to the simple network simulation model \obs{https://observablehq.com/@jwolondon/rampvis_idiom_odmap,1912}.
In multiple cases we found that:

\effective{New data spaces can be visualized informatively}

\subsubsection{Rapidly Refining and Relating Perspectives}
\label{sub:rapidlyrefine}

The ability to interactively change and modify data and views at speed is key to effective visualization
\obs{https://observablehq.com/d/880d234e47a1ed24,427}
(see also \cite{RSTA-2021-0309}).
It helps analyse data, refine problems and develop designs \obs{https://observablehq.com/@aidans/rampvis-idiom-gridded-glyphmaps,788} by selecting, combining and comparing items of interest; changing focus and scale; or varying the way that information is shown -- perhaps through alternative data spaces (\secref{sub:transform}).
Doing so in rapid and fluid fashion tightens design iterations and analytical work flow loops, which proved effective:
\sayIt{Interactivity was one of the most important aspects to me in terms of model development.
To be able to have a tool that I could interactively explore model outputs (and inputs) was very useful when adding new mechanisms or datasets} \obs{https://observablehq.com/@aidans/rampvis-idiom-gridded-glyphmaps,788}.

Interactivity was crucial to many of our successes - enabling modellers to see aspects of the model that would have been missed in static plots and adjust accordingly - e.g. \textit{Gridded GlyphMaps} for comparing (\secref{sub:variation}) two deterministic model runs \obs{https://observablehq.com/@aidans/rampvis-idiom-gridded-glyphmaps}, or when using parallel coordinate plots and heat maps to show parameter spaces \obs{https://observablehq.com/@henryqw/multiple-linked-views,400}, resulting in modellers being \sayIt{surprised by the amount of information made available by combining them} \obs{https://observablehq.com/@henryqw/multiple-linked-views,427}.


One important interaction involved rapid faceting and grouping that enabled us to explore variation in the data according to selected meta-data:
\sayIt{The ability to specify combinations of meta-data for grouping and sorting is effective to generate bespoke criteria aligned with the diverse needs of the application domain}. Doing so, through interactive visual comparison, led to insightful observations  \obs{https://observablehq.com/d/4aebb875cacaef3a,673}.
Indeed, \textit{Factoring in Flexibility} (\secref{sub:factorflexibility}) through rapid interaction was important throughout the collaboration: 

\effective{Rapid interaction underpins the processes that benefit from visualization}

\subsection{VISUALIZATION IS NOT ALWAYS APPROPRIATE}
\label{sub:visnotappropriate}

\summary{This is intended to say - you don't always have to use it, and failure does not discredit it or `disprove' the `theory' that it works. Just chose your engagements!} 


These considerable successes are a subset of those achieved and documented in the supplementary notebooks, which themselves constitute a sample of the total activity.
Yet the collaboration also showed that 
while graphical approaches can be applied to most data, they are neither necessary nor appropriate for all tasks, analysts or combinations of the two.
For example, good statistical techniques are more than adequate, for solving problems that are well defined with data that describe phenomena fully.
This is captured well by considering task clarity and information location 
\cite{sedlmair2012design},
\obs{https://observablehq.com/d/2e98f8d7f3cf5c08,832}.

But even where visualization offers potential, there may be good reasons for not pursuing a visualization engagement.
Data may not be forthcoming, collaborators may be unable to prioritise efforts to view, engage with or validate designs, and the kind of interpersonal relationships required for success may not be achieved.
Visualization is unlikely to be a priority, or a success, in such circumstances.
\textit{Design Study Methodology} \cite{sedlmair2012design} captures these \textit{pitfalls} well:
\activity{PF}{4}{no real data available (yet)};
\activity{PF}{5}{insufficient time available from potential collaborators};
\activity{PF}{11}{no rapport with collaborators}.
They are particularly likely to occur in the emergency response context.

Those offering visualization support must be sensitive to needs as they work to develop rapport, understand the problem domain, acquire data and establish whether a viable niche exists in which they can contribute: it may not.
Equally they must be able to move quickly as opportunities can open and close as the situation develops. Things may change: again highly likely during rapid response.
It's important to be engaged and available, but not disruptive,
to assess situations quickly, and to know when to move on.
Processes for visualization engagement and support need to be sensitive to these issues
for visualization to be used successfully across a complex project involving parallel efforts such as the SCRC volunteering.
As such, we do not identify individual examples in which visualization opportunities were not established here, but note that the reflection on constructs listed in Table~\ref{tab:suppMaterials} offer important insights, particularly through the \textit{What Did Not Work?} sections.

\subsection{VISUALIZATION IS EXPERIMENTAL}
\label{sub:experiments}

Just as visualization itself may not be required, so specific instances of it may be ineffective, despite apparent opportunity for transfer and good intention.
Existing approaches may not scale well to the problem or data in hand, or show the kinds of structures that are important. They may not be appropriate for computational reasons, perhaps as computational effort impedes the kind of interactivity required (\secref{sub:rapidlyrefine}).
An \textit{experimental} perspective frames the visualization design process as an opportunity to learn about the applicability and scope of visualization techniques --  under which conditions do particular idioms apply?
It encourages us to use design experiments to determine whether plausible visualization solutions `fit' across tasks and data sets, whether they scale and transfer to the new context and scope.
In line with an iterative take on the visualization design processes 
(e.g. \cite{sedlmair2012design, mccurdy2016action, hall2019design})
it encourages us to use design failures
to develop more appropriate and perhaps more transferable solutions.

Some of our well-intentioned efforts to transfer visualization techniques were unsuccessful design experiments, such as the use of colour for location: \obs{https://observablehq.com/d/43927395f6cb890c,596} and
offset lines for flow positions \obs{https://observablehq.com/@jwolondon/rampvis_idiom_directional_flow_curves,1453}.
Known approaches may not scale well, such as Hadlack\etals \textit{Balanced Representation Strategy} \cite{hadlak2011situ} for multiple linked view layout 
\obs{https://observablehq.com/@henryqw/multiple-linked-views,884}
or the use of animation to depict large numbers of changes \obs{https://observablehq.com/d/43927395f6cb890c,596}.
Pre-processing may be required in some cases due to data volumes \obs{https://observablehq.com/d/c3a628d8bd4fe8fe,733}.
Specific structure in some data sets made proposed designs ineffective, such as extreme variations in population density \obs{https://observablehq.com/@aidans/rampvis-idiom-gridded-glyphmaps,791}
or the sizes of the features revealed \obs{https://observablehq.com/@kaimdx/rampvis-idiom-gmap,705},
as did the size of some of the larger data sets.
Some designs are sensitive to low level design choices involving alternative layouts and orderings \obs{https://observablehq.com/d/880d234e47a1ed24,438}.
Design proposals may be deemed unsuitable for all sorts of complex and sometimes quite subtle reasons
\obs{https://observablehq.com/@ritsosp/rampvis-idioms-narrative-design-patterns,675}.
Sometimes technical limitations, and lack of time to achieve computational efficiency made speeds of computation and rendering inadequate for interactive response - a problem that is hard to fix quickly in the emergency context: \sayIt{given the time we had, the performance is not optimized} 
\obs{https://observablehq.com/@henryqw/multiple-linked-views,438}, a sentiment shared across our reflections (e.g. 
\obs{https://observablehq.com/@kaimdx/rampvis-idiom-gmap,709}, \obs{https://observablehq.com/@lborohfang/rampvis-idiom-integrated-algorithmic-tools-for-visual-ana,425}).
An experimental framing can stimulate useful creative thinking to redesign and modify approaches to meet new needs and apply them to new data sets with particular scales, structures or other characteristics.
Some of our insights (\secref{sec:success}) benefited from this approach.
%

Recording, reflecting and reporting \cite{rogers2020insights} on visualization design experiments `in the field' contributes beneficially to the discipline of visualization.
This is particularly the case in broad projects that require rapid response and involve parallel activity. 
Computational notebooks (\secref{sub:constructReflectionOn}) structured with prompts and strategies for design discourse have great potential in scaffolding this experimental process \cite{wood2018design,beecham2020design}.

\subsection{VISUALIZATION CAPABILITY IS COLLECTIVE} 
\label{sec:community}

The volunteering effort, and any insights or learning achieved from it, have been highly dependent upon the \textit{tacit knowledge, skills and experience} of a \textit{community of researchers}. The SCRC volunteers had varying levels of complimentary and overlapping expertise in visualization, epidemiological modelling and visualization for epidemiological modelling. 
They were able to work together to rapidly share perspectives, and select and apply a range of technologies and techniques from a broad landscape of disciplines and experiences.
The repertoire of potential solutions, and knowledge of how to combine, develop and apply them that enabled all of this, is of great value,
as is the ability to combine, develop and apply people who have existing loose connections through an active international research community.
Coming together at a time of crisis to assess cognitive and computational possibilities and how particular approaches and technologies are likely perform and combine in certain circumstances under unfamiliar conditions, demonstrates the importance of this knowledge and the strength of these connections.

The SCRC volunteering reveals that a \textit{community of capability} exists in the UK, and that in an emergency situation it can be operationalised to rapidly and usefully apply its knowledge, skills and experience to a broad range of problems across a disparate project.
With technologies that are mature, flexible, expressive and fast, and people who have the skills to use them quickly and effectively, this can be achieved directly and with immediate success \obs{https://observablehq.com/d/071ee158d5418d96,415},
\obs{https://observablehq.com/d/e7dc9a1ffd763d73,726}.
Volunteers came from academia and industry and offered expertise in various compatible development technologies, including
\textit{D3} \cite{bostock2011d3},
\textit{JavaScript} \obs{https://observablehq.com/d/071ee158d5418d96,841}, \textit{Java} \obs{https://observablehq.com/@aidans/rampvis-idiom-gridded-glyphmaps,941},
\textit{Processing} \cite{reas2003processing},
\textit{Python} \obs{https://observablehq.com/d/78b20aa4152547e2,676},
\textit{Jupyter} \obs{https://observablehq.com/d/43927395f6cb890c,787}
and
\textit{Observable} \cite{observable}.
Immediate successes can be extended through longer term creative problem solving and iteration, efforts that can be
informed by a theory-driven approach \obs{https://observablehq.com/d/087e459840b2b478}.
This is the case despite some instances across a project where visualization is not necessary or appropriate (\secref{sub:visnotappropriate}), and during an emergency in which individuals in a complex collaboration, and the connections between them, may be stretched, strained or broken.

\subsection{VISUALIZATION PROCESSES EVOLVE}
\label{sub:visprocess}

Establishing where and why and how visualization might apply most beneficially in complex new contexts is key to its effective use and core to visualization research.
Our collective experiences of COVID-19 visualization support through RAMP revealed plenty about the process of applying visualization knowledge in an emergency response.
Close interactions between visualization researchers engaging in \textit{Immersive Design} \cite{hall2019design} with the modelling teams, our central support for analytical processes and our efforts to disseminate information visually through a centralized architecture all contributed to this learning.
Just as design solutions apply and adapt known idioms, established design processes adapt and evolve when employed in different circumstances.
We therefore offer some preliminary ideas about the \textit{process} through which visualization design, visual analysis and visual problem solving may be conducted in collaborative rapid response projects. 

We do so by identifying \textit{10 characteristics of collaboration} that resonated in our reflection across the efforts.
These are somewhat interdependent, but characterise the kind of collaborative ecosystem of rapid experimental analysis and design that we found to be effective in supporting data analysts with visualization in emergency response.   
They were key to the successes reported in \secref{sec:success} and we suggest that they be adopted, assessed and adapted in future use of visualization in epidemiology and other applied settings.
We consider them to be recommendations for visualization designers and researchers that complement existing guidelines from the perspective of emergency response.

\subsubsectionALT{Rapid Response and Subsequent System}
{The Rapid Response Requirement\\(and Subsequent System Suggestion)}
{Respond Rapidly with Analytical Visual Interventions}
\label{sub:rapidresponse}

Conventional visualization support usually uses and encourages long-term collaboration \cite{shneiderman2006strategies, sedlmair2012design, meyer2012four, meyer2015nested, mccurdy2016action, meyer2019criteria} and results in system development (\secref{sub:visForEpMod}).
The end-goal is for data analysts to be provided with reliable systems that give them independent capability post-engagement.
Emergency response requires a different approach.
The focus of our modelling support teams was on getting viable reliable graphics of data made as quickly as possible \cite{lloyd2011human}, to understand data, demonstrate capability and assess needs.
Importantly, visual analysts were on hand to develop, use and interpret the bespoke graphics that they injected into the analytical discussion (\secref{sub:rapidlyrefine}), factoring in some flexibility (\secref{sub:factorflexibility}) through their skills and expertise.
Using tools and methods that are to hand to do so by rapidly generating and iterating on potentially useful graphics through analytical visual interventions is an important enabling first step
\obs{https://observablehq.com/d/7d339207ef90c483,757},
\obs{https://observablehq.com/@kaimdx/rampvis-idiom-gmap,728}
\obs{https://observablehq.com/d/2e98f8d7f3cf5c08,688}.
It can result
in rapid progress and the kind of mutual understanding of the problem, the data, each other and how visualization can help, that establishes vital capability, focus and trust.

A key aspect of this role is to scan possibilities quickly and \textit{Dig Deep} (\secref{sub:deepdig}) to search for high impact low cost solutions by using what is known and what is at hand.
Notions and guidance from \cite{lloyd2011human} and \cite{hall2019design} and the kind of loose integration described by \cite{matkovic2008interactive}  apply well in this context \obs{https://observablehq.com/@lborohfang/rampvis-idiom-integrated-algorithmic-tools-for-visual-ana}.
Responding rapidly in this way may lead to design ideas and shared knowledge that can be captured and applied subsequently, through further analytical interventions or in usable visualization systems for observation and dissemination as well as analysis \cite{chen2015may}.
But a usable UI and robust system is not an initial intention  \obs{https://observablehq.com/d/ea78b530cac2ee60,438}.
It likely requires additional time, resource, reflection and engagement \obs{https://observablehq.com/@aidans/rampvis-idiom-gridded-glyphmaps,788},
\obs{https://observablehq.com/@aidans/rampvis-idiom-gridded-glyphmaps,931}
and perhaps different underlying technology.

\subsubsectionALT{Factoring for Flexibility}
{The Flexibility Factor}
{Factor in Flexibility}
\label{sub:factorflexibility}

The ability to rapidly vary data, its resolution, aggregation and visual depiction is key to rapid response. It enables design, analysis, shared knowledge, trust, opportunity and data generation to progress in parallel, as a collaboration is supported by effective visualization:
\sayIt{The whole system is very flexible and allows us to do lots of different comparisons and visualizations, which has been really important with such a complex dataset in understanding what is going on} \obs{https://observablehq.com/@aidans/rampvis-idiom-gridded-glyphmaps,788}.

Intermittent and unpredictable levels of communication among collaborators during rapid response mean that building flexibility into design is essential:
\sayIt{lack of continuous access to epidemiologists throughout the process meant we had to build general / customizable solutions that could be quickly configured and demonstrated} \obs{https://observablehq.com/d/4aebb875cacaef3a,681}.

Flexible visualization designs get relevant graphics made rapidly.
They are robust to (and indeed initiators of) changes in interest, needs, focus and data. As such, technologies and techniques that support them are highly beneficial \obs{https://observablehq.com/d/596df309c41cca50},
\obs{https://observablehq.com/d/071ee158d5418d96, 671}
and processes that enable and encourage them are essential.
The concept of the `\textit{data sketch}' \cite{lloyd2011human} -- a \sayIt{loosely bounded collection of data, functionality and ideas} that scavenges from existing applications and develops through the design process -- is valuable here.
Computational notebooks that document designs (sec.~\ref{sub:documentdesign}) and implement them for analysis offer plenty of scope for developing data sketches, capturing and sharing knowledge and factoring in flexibility.
They promote discourse around design, analysis and re-use, and can be used for interactive design experiments
\obs{https://observablehq.com/@jwolondon/rampvis_idiom_directional_flow_curves,825}
\obs{https://observablehq.com/@aidans/rampvis-idiom-gridded-glyphmaps,918}
.

\subsubsectionALT{Analysis is Design and Design is Analysis}
{The Analysis is Design is Analysis Conundrum}
{Analyse through Design, Design through Analysis}
\label{sub:analysisisdesign}

It seems unhelpful to distinguish between \textit{analysis} and \textit{design} for analytical and model-developmental visualization in this context.
The two are inextricably linked and seemed increasingly inseparable in our experiences. This symbiosis may be useful in other contexts.

Our rapid analysis, involving injecting visualization into workflows, establishes needs and tests candidate designs.
For example, in our efforts to transform abstract data (sec.~\ref{sub:transform}) 
into two dimensions so that they could be seen and understood,
we were collectively forced to prioritise certain aspects of the data for analytical (and graphical) focus.
These priorities, in terms of tasks, aspects of data and analytical techniques, changed as the data structures were visually revealed
\obs{https://observablehq.com/d/78b20aa4152547e2,689}.
This is a tight and effective design/analysis loop - designs drive analysis, just as analysis drives designs
\obs{https://observablehq.com/d/2e98f8d7f3cf5c08,680}.
One way forward in this particular case is to fix a space into which data are projected (perhaps through transformation - \secref{sub:transform}), add data, add visual encodings, apply analytical techniques, interpret graphics and then creatively, analytically change view parameters.
If we have \textit{Factored in Flexibility} (\secref{sub:factorflexibility}), we can redesign in response to data in real time.
The result can be a revealing, iteratively developed, documented, symbiotic analytical design narrative \obs{https://observablehq.com/d/78b20aa4152547e2,689}.

\subsubsectionALT{Deep Dig}
{The Deep Dig Disposition}
{Dig Deep}
\label{sub:deepdig}

The rapidity of our response
took initial advantage of the deep, often tacit, knowledge in the visualization community (\secref{sec:community}).
Thinking back can be effective, and is perhaps quicker than searching sideways.
As individuals, visualization volunteers predominantly cast a deep rather than wide net - taking ideas that they had used successfully in the past, 
many of which they had been involved in developing,
as candidates for transfer, rather than those that were produced elsewhere.

For example, the architecture for generic support moved through Brodlie\etals five-level deployment model for visual computing \cite{brodlie2005visual,brodlie2007adaptive} with great speed due to the \sayIt{wonderful knowledge, skill, teamwork, and willingness to help combating COVID-19} \obs{https://observablehq.com/d/54c8641168c013ea,699} and the tried and tested nature of this useful theoretical construct. 

This emergency alternative to the costly \textit{Learn} phase of \textit{Design Study Methodology} \cite{sedlmair2012design} proved viable and effective given our rapid response emphasis (\secref{sub:rapidresponse}). 
The possible disadvantages of experience bias are countered by close collaboration with data experts through design discourse and an \textit{Experimental Approach} to visualization (\secref{sub:experiments}) in a \textit{Critical Creative Culture} (\secref{sub:creativeCulture}).
Self awareness and reflexivity can further moderate \cite{meyer2019criteria}.



\subsubsectionALT{Refine \& Combine}
{The Refine and Combine Protocol}
{Refine \& Combine to Design (and Analyse)}
\label{sub:refinceCombine}
\label{sub:refineCombine}

Refining and combining idioms is powerful, challenging and an essential part of any process of visual analysis and design \obs{https://observablehq.com/@jcrbrts/rampvis-idioms-fds,981}.
It requires visualization expertise, technical skill, close collaboration and open communication.
It is built on a \textit{Deep Dig} (\secref{sub:deepdig}) and helped (accelerated) by \textit{Factoring in Flexibility} (\secref{sub:factorflexibility}).
It can result in effective solutions, designed to meet specific challenging needs.

Our collaborations were full of examples including, among others,
our visual faceting work \obs{https://observablehq.com/d/4aebb875cacaef3a},
approaches that represent error and uncertainty \obs{https://observablehq.com/d/ea78b530cac2ee60},
linked parallel coordinates for visual parameter space exploration \obs{https://observablehq.com/d/7d339207ef90c483},
our gridded bi-directional flow maps \obs{https://observablehq.com/@jwolondon/rampvis_idiom_directional_flow_curves,1401}
and the development of an ontology-based visualization recommendation system \obs{https://observablehq.com/d/596df309c41cca50,746}.

\subsubsectionALT{Critical Creative Culture}
{The Culture of Critical Creativity}
{Create a Critical Creative Culture}
\label{sub:creativeculture}
\label{sub:creativeCulture}

A culture of critical creative thinking is an essential factor in achieving much of this, particularly in an emergency and in delivering the speed and flexibility that we advocate  (it is core to the ideas presented in sections ~\ref{sub:rapidresponse},~\ref{sub:factorflexibility},~\ref{sub:analysisisdesign}, \ref{sub:deepdig}  \& \ref{sub:refineCombine}).
It underlies the \textit{Experimental Approach} that we recommend (\secref{sub:experiments}).
The visualization knowledge base offers excellent proven methods, tools, functionality and guidelines that can be accessed quickly with a \textit{Deep Dig} (\secref{sub:deepdig}).
These can be used most effectively if applied and adapted creatively.
But high levels of openness, trust and communication are required to develop effective specific solutions to complex problems at speed.
The modelling support teams experienced this, feeling at times as though they were part of an \sayIt{idea generation lab} \obs{https://observablehq.com/d/78b20aa4152547e2,684}.

\begin{figure*}[htb]
  \centering
  \vspace{4px}
  \includegraphics[width=.99\textwidth]{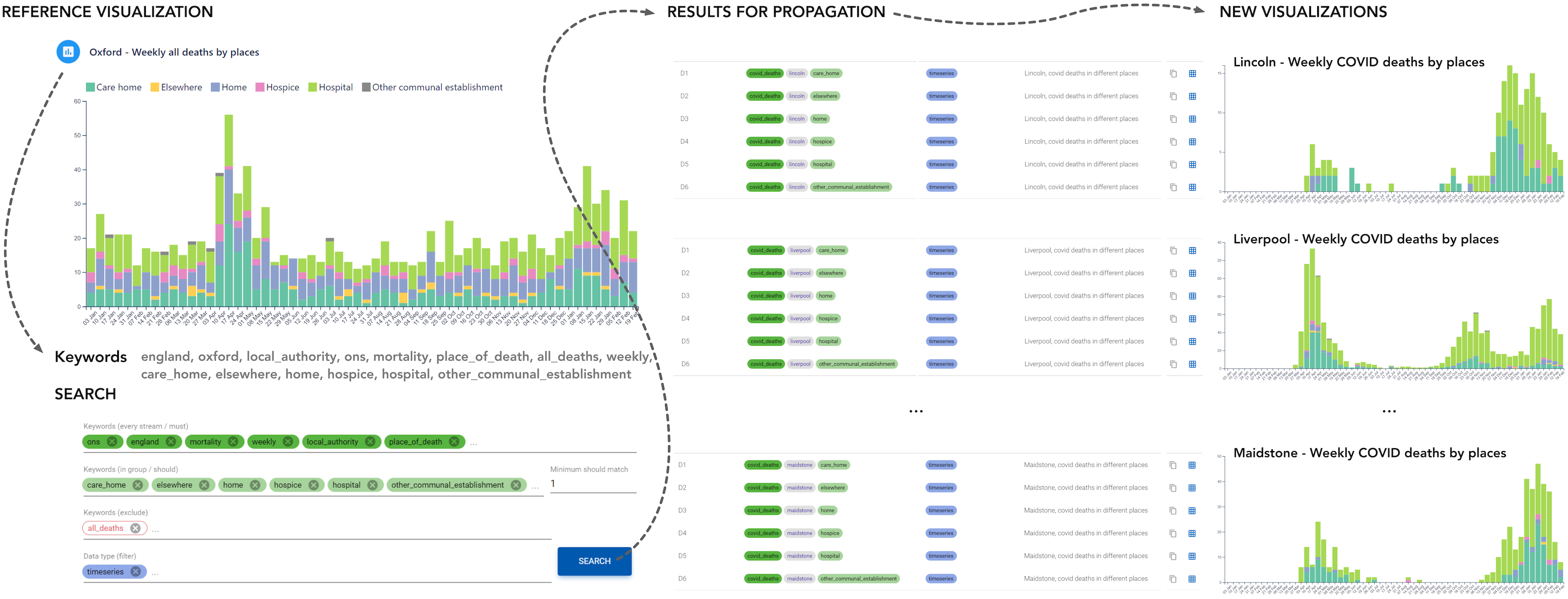}
  \caption{
  We developed an effective means of generating thousands of viable online interactive visualizations and hundreds of dashboards by developing a means of \textit{Propagating Visual Designs} from the SCRC data streams \cite{https://observablehq.com/d/87a416cd4468fff0}.
  This semi-automated process maps single or multiple data sets to particular visual forms (plot types).
  Our application also supports quality assurance over the propagation process, to ensure the propagated visual designs are reasonable.
  The approach uses an ontology \protect\obs{https://observablehq.com/d/596df309c41cca50} to facilitate propagation by formally representing the relationship between data set and visual device. This provides both flexibility and generalization as the mappings are determined by the ontology, which is established by human experts.
  The process provides a good example of the kind of human in the loop decision-making that visualization can support, resulting in down-stream efficiencies in this instance by encoding expertise that can be propagated and thus making good use of expert time.
  Here, visual analysis supports efficient and effective visualization design as we search for appropriate design solutions for \textit{observational visualization} \protect\obs{https://observablehq.com/d/a9aaed2f31718620,729}.
  }
  \label{fig:propagation}
\end{figure*}

In this kind of environment, constraints and lack of resources, or challenges such as a new scale of data or unexpected data structures, can drive creativity, innovation and collaboration (sec.~\ref{sub:experiments}).
Our approach to propagating visual designs (see Fig.~\ref{fig:propagation}) is a creative solution that enabled us to rapidly produce thousands of viable online interactive visualizations and hundreds of dashboards from limited volunteer developer resources \obs{https://observablehq.com/d/87a416cd4468fff0,695}.
Some inspired thinking enabled us to refine approaches and produce
new solutions that may in turn be of use elsewhere.
The population-proportional fading used to account for population densities in our gridded glyphMaps \obs{https://observablehq.com/@aidans/rampvis-idiom-gridded-glyphmaps,788} is a good example (see Fig.~\ref{fig:GGM}), as is the improved ontology for mapping data types to visual idioms
\obs{https://observablehq.com/d/596df309c41cca50,687}.
Equally, designing informative new spaces into which data are meaningfully depicted is an analytical/design process (\secref{sub:analysisisdesign}) that requires us to think deeply and creatively about complex analytical problems - both about what we want to know, and on the computational techniques and algorithms that we can deploy.
Here we ask: \sayIt{how can I compute new `data' that is effective, robust, and immediately relevant to the analytical task at hand?} \obs{https://observablehq.com/d/78b20aa4152547e2,689}. 


Specific methods can be applied to encourage creative thinking (e.g. \cite{goodwin2013creative, kerzner2019framework}).
The five design sheets approach \cite{roberts2016five} was useful in generating and refining ideas for disseminative visualization  \obs{https://observablehq.com/@jcrbrts/rampvis-idioms-fds}.
There is scope for further transfer and refinement of creativity methods for visualization as we establish a critical creative culture for design driven analysis and analysis driven design.
Such activity can potentially reduce the reliance on, and bias inherent in, the rapid response \textit{Deep Dig} (\secref{sub:deepdig}).

\subsubsectionALT{Documenting Design Discussion}
{The Design Discussion Document}
{Document Design Discourse}
\label{sub:documentdesign}

Open discourse around potential solutions that develop through this flexible creative approach to visualization and analysis can help move things forward as collaborators collectively learn about data, domains and designs.
Documenting this thoroughly is essential, to engage collaborators who are short of time (\secref{sub:factorflexibility}),
and record reasons and reactions \cite{beecham2020design}.
This may take time, but establishes a \sayIt{helpful compromise} \obs{https://observablehq.com/@aidans/rampvis-idiom-gridded-glyphmaps,788} that provides pertinent, interpretable visual information, in line with established good practice (e.g. \activity{S}{2}{Document} \cite{hall2019design}) \obs{https://observablehq.com/d/2e98f8d7f3cf5c08,688}.


The aims are various:
to diminish knowledge boundaries between visualizers and domain specialists through knowledge sharing \obs{https://observablehq.com/d/2e98f8d7f3cf5c08,680} -- (flexible) visual artefacts from both sides of the knowledge divide can be useful boundary objects \cite{arias2000transcending} to help share ideas and perspectives  
and to inform  discourse that leads to design ideas and analysis
\obs{https://observablehq.com/@jsndyks/rampvis-idiom-design-by-immersion,910};
to develop a fully documented, analytical design narrative that explains progress with examples and rationale that encourage scrutiny \obs{https://observablehq.com/d/2e98f8d7f3cf5c08,688} \cite{meyer2019criteria};
to establish a record that can be shared with new collaborators, across projects or as the basis for post-project reflection and learning; etc. 



\subsubsectionALT{Synchronise \& De-synchronise}
{The Synchronise De-synchronise Set-up}
{Synchronise \& De-synchronise}
\label{sub:synch}
\label{sub:cadence}

Establishing a cadence and a commitment to communication is a particular challenge in uncertain and busy times. Understandably, those providing visualization support are sometimes working in the dark, and need to establish ways of getting efficient and effective feedback that informs and does not disrupt. A mix of regular short synchronous meetings (cadence)
\obs{https://observablehq.com/d/7d339207ef90c483,966}
with a reasonable and negotiable attendance target (commitment) -- \sayIt{quick, agile cycles} \obs{https://observablehq.com/d/2e98f8d7f3cf5c08,680} -- with documented descriptions that provided scope for asynchronous feedback
(with deadlines) offered an effective balance between flexibility and commitment in many cases.
This was particularly so when supported by good documentation and searchable asynchronous digital communications streams (\secref{sub:documentdesign}).
%
Regular logging \obs{https://observablehq.com/@aidans/rampvis-idiom-gridded-glyphmaps,811} and transparent language \obs{https://observablehq.com/d/2e98f8d7f3cf5c08,688} help capture progress and keep things moving when schedules, levels of commitment and priorities are dynamic and unpredictable.

Visualization volunteers found it beneficial to get `in sync' with the modeller’s processes where possible
\obs{https://observablehq.com/d/efe6c1c90c625bec,723}
but found it important to prepare for and accept down time, and make use of asynchronous methods too. Asynchronous activity works both ways and can involve:
\begin{itemize}
  \setlength\itemsep{0pt}
\item \textbf{direction} -- asking collaborators for resources through which knowledge can be built offline - papers, examples, data sets, project documentation, then
\item \textbf{connection} -- using rapidly developed visualization as a common boundary object \cite{arias2000transcending} to re-connect either synchronously at planned meetings
\obs{https://observablehq.com/@henryqw/multiple-linked-views,877}
or asynchronously through materials that document discussion around design and analysis (e.g. \cite{beecham2020design}).
\end{itemize}
One modelling support team found it useful to characterise this as a \sayIt{hurry up and wait} strategy
\obs{https://observablehq.com/d/a9aaed2f31718620,725}.

There are open challenges around scaling this process and making it efficient - teams grow and shrink and team members priorities and availability change rapidly in an emergency.
But we found it a useful set-up in circumstances that differed from more regular visualization design engagements. It may transfer well to non-emergency contexts (see \cite{beecham2020design}), but is very dependent upon \textit{Documented Design Discourse} (\secref{sub:documentdesign}) and \textit{Factoring in Flexibility} (\secref{sub:factorflexibility}).

\subsubsectionALT{Lurk to Learn}
{The Lurk to Learn Tactic}
{Lurk to Learn}
\label{sub:lurk}

Participating in video conference calls in listening-only mode, with permission, can be an excellent non-disruptive means of learning.
This is a complex process as priorities and roles are dynamic.
We found lurking and listening to work exceptionally well, have little cost and be particularly unobtrusive in the digital workspace.
It proved useful for improving understanding, identifying roles for visualization and selecting candidate data sets and problems to which
visualization design might be usefully applied:
\sayIt{Lurking in these meetings may provide a low-stress way to get ideas for visualization tasks or users, especially when the case for visualization is not clear to those involved.}
\obs{https://observablehq.com/d/a9aaed2f31718620,725}.

Subsequently dropping candidate designs into  discussions that address these problems can help communicate possibilities, stimulate discourse around data that establish roles and priorities, and hopefully contribute to data interpretation and analysis.
This is a great role for lurking (friendly) \sayIt{visualization spies} \obs{https://observablehq.com/d/7d339207ef90c483,966}.
Perhaps, the post-COVID-lockdown, rapid response visualization researcher/designer/analyst immerses \cite{hall2019design} by \textit{lurking} in digital workspaces and meetings?
It was very effective in many cases in our project, 
enabling us to \activity{S}{1}{Observe domain experts practices unobtrusively in situ}.
This provided a vital and unexpected means of learning, assessing possibilities and rapidly and effectively conducting design and analysis experiments by injecting visualization into workflows where it might be beneficial \obs{https://observablehq.com/d/2e98f8d7f3cf5c08,680}, \obs{https://observablehq.com/@jwolondon/rampvis_idiom_directional_flow_curves,1447}. 

\subsubsectionALT{Embrace Digital Immersion \& Communication}
{The Digital Immersion Opportunity}
{Embrace Digital Immersion}
\label{sub:digital}

Indeed, the pandemic, its various lockdowns, and the associated rapid emergence, acceptance and shaping of screen sharing video meetings, with broad bandwidth and high resolution imagery, have resulted in a rich digital workspace for analysis and collaboration \obs{https://observablehq.com/@jwolondon/rampvis_idiom_odmap,2529}.
This suits visualization and visual analysis well and many of these themes that characterise our experience of applying visualization process to support epidemiological modelling are enabled by this technology. This includes activity that is paper-based \obs{https://observablehq.com/@jcrbrts/rampvis-idioms-fds,821}. It equips us to build on and improve existing approaches to developing engagement between visualization researchers and domain experts, as described variously above.  

The objective to \sayIt{Embed VIS expertise in the workflow} \obs{https://observablehq.com/d/2e98f8d7f3cf5c08,688} seems an important first step and good precursor to the system building that is typical of applied visualization research. The digital workspace undoubtedly helps with this.
But it also enables us to continue throughout a collaboration as we move from \textit{Rapid Response} (\secref{sub:rapidresponse}) to something more permanent in ways that that may not involve the post-engagement isolation assumed by system building approaches that pre-date the digital workspace.
How many visualization system user guides say \sayIt{just call me on Zoom or ping me on Teams and we can work on this together through screenshare}?
This frequently became our mode of providing visualization support in the digital collaboration of SCRC, much of which was undertaken during lockdown when people were unable to leave their homes.
Perhaps a situation in which expert analysts work independently with stable software is less necessary or even appropriate in post-pandemic digital workspaces?
Perhaps in this context the interdependence of concurrent \textit{Design through Analysis through Design} (\secref{sub:analysisisdesign}) requires some new thinking about the relationships between visual analyst, data analyst and technology and the roles that support them in the medium and long term.
We wonder what happens when we combine \textit{Design by Immersion} \cite{hall2019design}
with GitHub, notebooks, Zoom and collaboration tools like Slack,
Figma (\obs{https://observablehq.com/d/071ee158d5418d96,852}) and
Zulip (\obs{https://observablehq.com/@jsndyks/rampvis-idiom-design-by-immersion,933})?
It may be informative and transformative to find out.
Notions from other applied visualization design methodologies such as
\textit{multi-dimensional in-depth long-term case studies} \cite{shneiderman2006strategies},
\textit{multi-channel visualization engagements} \cite{wood2014moving} and \textit{action design research} \cite{mccurdy2016action} may help us embrace the digital workspace in immersive collaborative analytical visualization design.

\subsection{VISUALIZATION INSPIRES RESEARCH}

Our cross-study analysis shows that the processes and roles and activities associated with applied visualization design are changing. Section \ref{sub:visprocess} offers some insights. But it also reveals some open problems that were either reported in multiple notebooks or that seem relevant to multiple activities across the SCRC engagements.
These findings demonstrate the important role of applied work in inspiring and stimulating visualization research. 
They should be addressed through a combination of applied visualization work in epidemiology and elsewhere, and controlled experimental approaches that complement this activity.
In short, we need to establish, develop, assess and refine...

\begin{itemize}
  \setlength\itemsep{0pt}
   \item effective and efficient ways of \textit{visualizing differences} in quantities that allow us to make spatial and temporal comparisons \obs{https://observablehq.com/@aidans/rampvis-idiom-gridded-glyphmaps,788} (see also \cite{RSTA-2022-0039} \textit{Fig 2.} for a temporal view of a spatio-temporal data set); 
   \item methods for comparing quantities and ratios that  \textit{vary by orders of magnitude} \obs{https://observablehq.com/@jwolondon/rampvis_idiom_odmap,2380} (see also \cite{RSTA-2021-0298}) and that \textit{control for population size} \obs{https://observablehq.com/@aidans/rampvis-idiom-gridded-glyphmaps,788};
   \item effective ways of using \textit{layout and colour in combination} in dense data graphics \obs{https://observablehq.com/d/880d234e47a1ed24} (see also \cite{RSTA-2022-0039} \textit{Fig 4.});
   
   \item visualization idioms to deal with \textit{large numbers of data items} and \textit{new structures in data} that are unexpected or important
   \obs{https://observablehq.com/@kaimdx/rampvis-idiom-gmap,705},
   \obs{https://observablehq.com/d/596df309c41cca50,687};
   
   \item \textit{consistent visual languages} - something that is hard to achieve in a pandemic (in parallel) - that allow us to use colour, icons, other encodings and even interactions in ways that are common, predictable, consistent, effective and understood \obs{https://observablehq.com/@ritsosp/rampvis-idioms-pictograms}; 
   \item \textit{narrative patterns} for communicating in cases where subjects are sensitive or controversial \obs{https://observablehq.com/@ritsosp/rampvis-idioms-narrative-design-patterns}, \obs{https://observablehq.com/@ritsosp/rampvis-idioms-pictograms,706};
   
   \item approaches that \textit{minimise misinterpretation} and account for it where it occurs  \obs{https://observablehq.com/@aidans/rampvis-idiom-gridded-glyphmaps,788} by addressing some of the open issues listed above, and through effective documentation, signposting, training, learning and co-design processes;
   \item effective ways of further \textit{embracing the emerging digital workspace} for long-term immersive visualization support;
   \item reliable and effective \textit{processes for conducting and supporting research through applied visualization} that draw upon the themes identified through this engagement between epidemiological modellers and visualization researchers (\secref{sub:visprocess}). 
\end{itemize}

These open issues demonstrate ways in which deep engagement in data rich problem domains infuses visualization research with potentially impactful opportunities and inspiring challenges.


\section{Recommendations}
\label{sec:recommendations}

These experiences and the claims, ideas and findings that they have enabled us to develop,
lead us to 
some \textit{tentative recommendations}.
They address the visualization research community, the epidemiological modelling community, the wider scientific community and the bodies that support and sustain the activities of these groups.
Just like the claims and themes that we document, and the parallel projects that have given rise to them, they are
highly interdependent. 

\subsection{Reward and Invest in Reusable Functionality}
\label{rec:reward}

Research that involves working implementations of functionality, with full descriptions and usable code is hugely valuable.
It allows visualization knowledge to be rapidly and successfully applied to and tested in important new contexts, effectively and with efficiency.
It is key to many of the characteristics of collaboration that we identify and should be fully supported. 
Well documented accessible libraries of functionality help with transfer, reliability, creativity and speed
\obs{https://observablehq.com/d/78b20aa4152547e2,925},
\obs{https://observablehq.com/d/4aebb875cacaef3a,681}.
They enabled much of what we achieved, underpin many of our findings and are fundamental to much of our guidance.

We need to \textit{develop and maintain usable, documented, open libraries of VIS functionality}.
These are somewhat undervalued outputs in the academic environment and often regarded as an expensive add on to academic contributions.
Where they work, they can be enormously effective and influential \cite{bostock2011d3, lex2014upset, conway2017upsetr}.
Without them, activity is slowed down by re-implementation
\obs{https://observablehq.com/d/4aebb875cacaef3a,720} and
high impact interventions may not be possible \obs{https://observablehq.com/d/2e98f8d7f3cf5c08,684}.
%
They could be supported by efforts to improve reusability \obs{https://observablehq.com/d/071ee158d5418d96,710} and would enable the kind of agile and flexible response advocated here \obs{https://observablehq.com/d/e7dc9a1ffd763d73,722}.
We would like to see such core contributions funded, coordinated, appreciated and rewarded in anticipation of future projects that involve rapid response to unknown emergencies and the less urgent opportunities to adopt and adapt visualization methods to a variety of ongoing problems.
The \textit{Observable} notebooks presented here \cite{dykes_zenodo5717367}, and the RAMP VIS project outputs, are a small part of this process.

\rec{We encourage the research community and its funders to invest in and deliver research that empowers by developing and delivering open reusable visualization and visual analytics functionality} 

\subsection{Value and Use Applied Visualization Design Research}
\label{rec:applied}
We found out plenty through our applied visualization experiments, and moved the body of knowledge forward somewhat.
But we still lack \sayIt{specific guidance about what visualization techniques and interactions will be most effective for which mid-level tasks.} \obs{https://observablehq.com/d/7d339207ef90c483,713}.

Colleagues report, for example, that \sayIt{The [visual faceting] idiom seems highly suited to COVID related analyses due to the complexities of the data points themselves and their limited number. On the negative side the idiom is somewhat poorly explored in existing literature and there isn't sufficient guidance and support for its deployment} \obs{https://observablehq.com/d/4aebb875cacaef3a,669}.
To make the most of these opportunities for VIS we need to \sayIt{explore the use of the idiom in real analysis settings - how is it used, what analyses does it support, what are the scenarios that it maps to best?} \obs{https://observablehq.com/d/4aebb875cacaef3a,681}. 

We recommend a focus on applied VIS research that will advance our ability to map techniques to data and tasks, to build the visualization knowledge base through experience.
Exploratory experimental visualization work in deeply applied contexts will draw attention to open problems in visualization and help us understand the potential of graphics in discovery beyond epidemiology and emergency response.

\rec{We encourage the research community and its funders to engage in a broad programme of applied visualization research to explore and develop visualization knowledge in wide contexts}

\subsection{Document in Detail to Learn and Share Knowledge}
\label{rec:document}
Clear and comprehensive communication of applied visualization research, involving detailed descriptions of context, with reflective commentary and discussion, is important for knowledge-sharing.
It offers potential for transferring knowledge \textit{between contexts} for application in rapid response and elsewhere.
It also provides opportunity for cross-study meta-analysis that draws upon diverse experiences that we see so rarely in visualization.
This is particularly true if claims and contexts are structured, searchable and findable. 
This paper and its supplementary digital notebooks offer an example of how this might be achieved.

Reporting in this way is also important \textit{within} diverse projects such as the RAMP efforts to use visualization in epidemiological modelling. It enables those working in them to share ideas and experiences, supports asynchronous work, justifies decisions, makes alternatives explicit and may contribute to the validity of the work undertaken \cite{meyer2019criteria}.
It is fundamental to many of the processes of analysis and design that we recommend.
Our efforts to do so enabled us to find common activities such as similar task requirements that occur across different modelling workflows \cite{https://observablehq.com/d/a9aaed2f31718620}, share and broaden solutions (e.g. developing consistent colour schemes and shared analytical functionality across the project), and identify cases that required a re-think (e.g. the various compartmental models that required comparable visual treatment).

Producing informative documentation is time consuming but rewarding.
The unpredictable and variable cadence of the emergency effort made it both crucial and (with planning and flexibility) feasible during our volunteering.
While our emergency response work was something of a unique opportunity for this kind of activity the possibilities are broader: for communication between groups, individuals, topics, funding rounds, etc.
The current culture of reward structures and systems does not particularly encourage this kind of detailed documentation, but the notebook paradigm (see also \cite{wood2018design}) and our adoption of it in the visualization context have provided a means for sharing across the group, prompting reflection and establishing themes with some success. This model could be applied elsewhere for communication and analysis within and between applied visualization projects.

\rec{We encourage the research community to develop a culture of comprehensive structured reporting within and between applied visualization projects, and to utilise documented experiences from a range of projects to develop reliable knowledge and inform approaches}

\subsection{Sustain and Strengthen the Visualization Community}
\label{rec:sustain}

The volunteering activity constitutes valuable progress in terms of the connections, cooperation, capability, knowledge and good will that has resulted within and between the epidemiological modelling and visualization communities.
In many ways, what we have done has also shaped the visualization community well for future support efforts in the UK.
This may involve wider scientific endeavour, future emergency response and the very prescient challenges facing society that involve large, complex, rich data sets.
Understandably, we would like to push forwards and build upon what has been achieved, rather than lose it.

We have established important skills, talents, software, infrastructure, teams, connections, and ways of working  while developing knowledge and capacity.
Only some of these are captured in documents such as this one.
%
Much of this rich ecosystem of capability involves tacit knowledge that must be shared, sustained and strengthened through investment in people: 
\sayIt{We should not let this effort, the resultant software, and perhaps most importantly the accumulated R\&D knowledge go to waste ...
The underlying infrastructure that was generated by the RAMP VIS team should be consolidated and made available for future use.}
\obs{https://observablehq.com/d/54c8641168c013ea,705}

Given this, it seems important and wise to maintain, support, develop and apply the body of knowledge, collective capability and expertise that makes this possible.
This may involve supporting researchers to work in new areas and to develop their knowledge and skillsets (e.g. UKRI, EP/V033670/1, \textit{Visualising Contact Networks in Response to COVID-19)}
and increasing Research Software Engineer resources for transforming designs and prototypes to the VIS infrastructure (e.g. UKRI EP/V054236/1, \textit{RAMP VIS}).
But this might be best supported by coordinated strategic investment.  

\rec{We call for investment in people to sustain and strengthen the visualization community, to leverage and consolidate the progress made 
and apply what has been learned to challenges and domains beyond epidemiological modelling}


\subsection{Adopt Visualization Design for Analysis}
\label{rec:analyticDesign}
We found rapid iterative design experiments to work well in capturing analytical requirements, showcasing possibilities, learning visual languages and prioritizing activity as we
apply VIS techniques to complex data challenges in new configurations.
They also contribute to the team building, knowledge sharing and development of common understanding that underpin such efforts, through actual analysis:
\sayIt{The requirement and design work seem to become part of the visualisation creation} \obs{https://observablehq.com/@kaimdx/rampvis-idiom-gmap}.
%
We should not be too surprised by this (see \cite{lloyd2011human}).
But the roles of visualization design in analytical steering and trust building are not adequately acknowledged in the dominant software system development process models that describe applied visualization research \cite{sedlmair2012design}.
Equally the wider scientific community may not be fully aware of the potential for VIS in all stages of the scientific process: in model building and analysis as well as for observation and dissemination.
VIS is a human analytical process rather than a technology. 

To address this, visualization researchers  should embrace the perspective of the rapid \textit{Visualization Design Experiment} (\secref{sub:experiments})
to be less focused on system development.
This will enable a sharper focus on the inseparable relationship of discovery between analysis and design (\secref{sub:rapidresponse}, \secref{sub:analysisisdesign}) that enables us to assess and refine opportunities for VIS and developing domain knowledge iteratively and experimentally (\secref{sub:experiments}).
It may help us use visualization more widely in the analytical stages of scientific enquiry.
Developing a creative culture (\secref{sub:creativeculture}) that allows us to build knowledge through applied analytical design work is key to this.

\rec{We encourage the scientific community to  embrace analytical visualization and recommend that the visualization research community develops its approaches to applied visualization activity to account for
the inseparability of the design and analysis processes}


\section{Conclusions}
\label{sec:conclusion}
\summary{A short section in which we say something about what all of this means and how we need to take things forward. This should be a call to action and will need careful consideration. This is the place to emphasize our core message, and should be supported by the evidence and narrative that precedes it.
We did plenty and have learned plenty. The work is ongoing. We urge those using data to problem solve, to embrace visualization, visualization research and the visualization research \& design community to collaborate in ways that use visualization beneficially to develop, assess and communicate knowledge in critical domains. We hope that these recommendations and the resources provided here and elsewhere in RAMPVIS will be persuasive and will help with this process. 
}

This paper reports on a complex project, involving parallel activity across a broad range of efforts to provide visualization support for epidemiological modelling.
It presents a diverse body of supplementary materials that document this work and capture the iterative structured thinking that has taken place around it \cite{dykes_zenodo5717367}.
It draws widely upon this data source to propose findings and develop recommendations.
The findings are preliminary and partial, but 
draw upon the knowledge, experience, collaborative capabilities and thinking of visualization researchers, epidemiological modellers, software engineers and other volunteers. 
The intention is to 
  \textit{develop knowledge about visualization in epidemiology that can be applied to other contexts}
  while providing examples that \textit{demonstrate how this might be achieved}
  so that \textit{visualization can be used effectively and widely in data-driven research}.



We find plenty of opportunity for effective visualization to support epidemiological modelling at all four levels as we
\textit{observe}, 
\textit{analyse}, 
\textit{model} and 
\textit{disseminate}
data and its derivatives.
We also find calls for more interactive VIS tools for exploration and provenance elsewhere in this special issue \cite{RSTA-2021-0300}.
Visualization is not guaranteed to be effective, but applying visualization knowledge and adapting visual analytic approaches in a creative experimental environment of collaborative data analysis and design can be informative. 
Approaches that involve design by immersion \cite{hall2019design} help in the early stages of intense collaboration, where the focus is more on analysis and knowledge sharing than system design.
Subsequent software systems may provide proven persistent functionality and benefit from these early interactions, but extending the initial emphasis on knowledge sharing and trust building achieved through rapid analysis and insight marks a shift in current visualization design study thinking \cite{sedlmair2012design}.
This has been jolted by our emergency context,
supported by the digital workspace and informed by our visceral volunteering experiences.
Changes to visualization support processes that account for this refocus are helpful, including a greater emphasis on the deep symbiosis between design and analysis, and the value of early activities in sharing knowledge, understanding data, refining tasks and establishing needs.
The recommendations made in section~\ref{sec:recommendations} build upon these findings in an effort to maintain momentum, make the most of the knowledge achieved and support the VIS and wider scientific communities in developing and applying visualization knowledge.
Acting upon them is highly likely to deliver many more successes, such as those listed in section~\ref{sec:success}.


Importantly, the rich
series of interactive reflective notebooks that we offer to supplement this contribution acts as a considerable evidence base to support these claims.
This `data' documents selected examples of visualization knowledge and comments on its use in the RAMP activity.
Providing it enables us to refine, share and discuss our experiences and interpretations of them as knowledge settles and we continue to learn through iteration, reflection and critique.
We hope that the collection of interactive notebooks supports our claims well, and in ways that allow others to scrutinise and evaluate them. 
We see this form of rich documentation as important in the processes of recording, reflecting upon and reporting applied visualization research.
The notebooks demonstrate how visualization knowledge was transferred to, shaped in and developed by the particular problem in hand.
In some cases they provide working functionality that can be reused and applied elsewhere.
We hope that they show how VIS might be applied and developed in other contexts across science - directly with examples, and indirectly through rich description of context and subsequent inference.


\acknowledgments{

This work was undertaken in part as a contribution to the Rapid Assistance in Modelling the Pandemic (RAMP) initiative, coordinated by the Royal Society.

The authors would like to express their gratitude to the SCRC management team for its leadership, all SCRC members for their selfless efforts, the UK Science and Technology Facilities Council (STFC) for providing hardware to SCRC and RAMP VIS, the STFC for providing and managing the SCRC hardware platform and the Royal Society for coordinating the initiative.

Our thanks go especially to the many modelling scientists and epidemiologists who worked with the VIS teams and provided valuable expert advice as well as datasets to be visualized.
They include Dr. Dexter Canoy, Dr. Glenn Marion, Dr. Iain McKendrick, Dr. Christopher Pooley, and Dr. Thibaud Porphyre.

In particular, we would like to thank all other VIS volunteers for their contributions to various VIS activities during 2020, including
Dr. Helen C. Purchase, Dr. Stella Mazeri, Dr. William Teahan,
Fedor Gorokhovik, 
Benjamin Nash, Tianci Wen and James Scott-Brown.

We would like to express our appreciation to
Alys Brett for guiding us in producing open-source software,
Sonia Mitchell for offering innumerable pieces of advice on data products,
Julie Meikle for advising on public engagement, and
Judy Dendy, Graeme Smith, and Steven Williams for their administrative support to the RAMP VIS project.



\section*{Funding}

This work was supported in part by the UKRI/EPSRC grants EP/V054236/1 and EP/V033670/1 and UKRI/STFC grant ST/V006126/1.

\section*{Author Affiliations}

\setlength{\leftskip}{0.5cm}

$^{1}$City, University of London, UK\\
$^{2}$King's College London, UK\\
$^{3}$Swansea University, UK\\
$^{4}$University of Edinburgh, UK\\
$^{5}$University of Oxford, UK\\
$^{6}$Loughborough University, UK\\
$^{7}$Nottingham University, UK\\
$^{8}$University of Glasgow, UK\\
$^{9}$Biomathematics and Statistics Scotland, UK\\
$^{10}$University of Chester, UK\\
$^{11}$UKAEA, UK\\
$^{12}$Bangor University, UK\\
$^{13}$Warwick University, UK\\
$^{14}$University of Sheffield, UK\\
$^{15}$Middlesex University London, UK

\setlength{\leftskip}{0cm}

}

\bibliographystyle{abbrv-doi}


\bibliography{bib/rampvis}

\bibliographystyleobs{abbrv}
\bibliographyobs{bib/notebooks}

\end{document}